\documentclass[a4paper,fleqn,usenatbib]{mnras}

\usepackage{newtxtext,newtxmath}
\usepackage[T1]{fontenc}
\usepackage{ae,aecompl}

\usepackage{graphicx}	% Including figure files
\usepackage{amsmath}	% Advanced maths commands
\usepackage{amssymb}	% Extra maths symbols

\title[Luminosity outburst chemistry]{Luminosity outburst chemistry in protoplanetary discs: going beyond standard tracers}

\author[D. S. Wiebe et al.]{
Dmitri S. Wiebe,$^{1}$\thanks{E-mail: dwiebe@inasan.ru}
Tamara S. Molyarova,$^{1}$
Vitaly V. Akimkin,$^{1}$ \newauthor
Eduard I. Vorobyov,$^{2,3}$ 
and Dmitry A. Semenov$^{4,5}$
\\
$^{1}$Institute of Astronomy, Russian Academy of Sciences, Moscow, 119017, Russia\\
$^{2}$Department of Astrophysics, University of Vienna, Vienna 1180, Austria \\
$^{3}$Research Institute of Physics, Southern Federal University, Stachki Ave. 194, 344090, Rostov-on-Don, Russia\\
$^{4}$Chemistry Department, Ludwig Maximilian University, Butenandtstr. 5-13, D-81377 Munich, Germany\\
$^{5}$Max Planck Institute for Astronomy, K\"onigstuhl 17, 69117, Heidelberg, Germany
}

\date{Accepted XXX. Received YYY; in original form ZZZ}

\pubyear{2017}

\begin{document}
\label{firstpage}
\pagerange{\pageref{firstpage}--\pageref{lastpage}}
\maketitle

\begin{abstract}
The chemical influence of luminosity outbursts on the environments of young solar-type stars is explored. Species are categorised into several types according to their response to the outburst. The first and second types imply chemical changes only during the outburst (with slightly different behaviours). These response types are mostly observed close to the star and are caused by icy mantle evaporation. However, mantles recover after the outburst almost immediately. A notable exception is benzene ice, which is accumulated on dust surfaces during and after the outburst, so that its abundance exceeds the pre-outburst level by orders of magnitude. The third type of response is mostly seen at the disc periphery and implies alteration of abundances during the outburst and preservation of these `abnormal' abundances for centuries. This behaviour is typical of organic compounds, like HCOOCH$_3$, CH$_3$CN, CH$_2$CO. Their presence in the dark disc regions can be a manifestation of the past outburst. CO and CO$_2$ only trace past outbursts at the remote disc regions. The outburst changes the C/O ratio, but it quickly returns to the pre-outburst value almost everywhere in the disc. An important factor determining the sensitivity of molecular composition to the outburst is the dust size distribution. The duration of the pre-outburst stage and of the outburst itself influence the chemical effects, if the burst duration is shorter than 50 yr and the duration of the quiescent phase between the bursts is shorter than 100 kyr.
\end{abstract}


%%%%%%%%%%%%%%%%%%%%%%%%%%%%%%%%%%%%%%%%%%%%%%%%%%

%%%%%%%%%%%%%%%%% BODY OF PAPER %%%%%%%%%%%%%%%%%%

\section{Introduction}

It now becomes increasingly evident that luminosity outbursts represent an inevitable and vital aspect of an early evolution of a low-mass protostar surrounded by a circumstellar accretion disc \cite[for a review, see][]{audard}. The evidence is also growing that luminosity outbursts play an important role in the evolution of massive (proto)stars in the primordial and present-day Universe \citep{Vorobyov2013,Hosokawa2016,Sakurai2016,2017MNRAS.464L..90M}.

The idea that FU Orionis-type luminosity outbursts (hereafter, FUors) accompany the early stellar evolution and are related to rapid protostellar accretion dates back to \cite{Herbig1966,Herbig1977} papers and to a note by \cite{1986PASP...98.1103K}. Since then, the idea has been supported by observations, and various numerical models have emerged including the thermal instability \citep{Bell1994}, disc gravitational fragmentation followed by infall of gaseous clumps \citep{VB2006,2015ApJ...805..115V,2017MNRAS.464L..90M,Zhao2018}, magnetorotational instability  \citep{Armitage2001,Zhu2009}, disc perturbation by a close stellar companion or by a (proto-)giant planet \citep{Bonnell1992,Pfalzner2008,Nayakshin2012}, etc.

The exact mechanism or mechanisms triggering the outburst are still not fully understood, because most known FUors have only been observed in detail in the high state, once the luminosity outburst has commenced. A notable exception is V346 Nor, which has shown a significant decrease in the accretion rate prior to a new outburst in 2011 \citep{norma,Kospal2017}. While stars experiencing FUor events are rare, numerical models indicate that the time interval between outbursts exceeds the duration of an individual outburst by a wide margin, hence outburst star+disc systems could actually be much more numerous \citep{Sholz2013,2015ApJ...805..115V}.

To make better statistical estimates of the FUor frequency, one needs to know observables that would allow revealing an outbursting star long after the end of the actual outburst. One of the most promising tracers of the past outburst activity is the disc chemistry. It was recognised more than a decade ago that the outburst may leave a long-lasting mark in the abundances of certain chemical species \citep{lee2007}.

Indeed, the disc chemical composition depends strongly on its temperature and on the impinging radiation. It must be kept in mind that even though the UV and optical radiation from the star and from the accreting region cannot penetrate deep into the disc, it is still capable of heating the disc midplane due to re-radiation of the absorbed energy in the infrared band. Thus, a luminosity outburst heats up the midplane and raises the UV-illumination in the outer disc regions. These factors stimulate icy mantle evaporation, enriching the gas phase by products of surface reactions, and may also trigger some high temperature gas-phase reactions \citep{2016ApJ...821...46T}, similar to those that are expected in hot cores \citep{2013ChRv..113.8939G}. The combined effect of the outburst-related changes in the physical conditions may not fade away immediately after the end of the outburst, so that we may be able to detect traces of past outbursts by observing `non-typical' abundances of certain molecules. In addition, study of molecular composition during the outburst and post-outburst phases opens a window into the surface chemistry that have taken place during the quiescent phase.

Attempts to find reliable past outburst tracers in discs and surrounding envelopes have been so far mostly focussed on simple molecules like CO and CO$_2$. The essence of the expected chemical changes is the CO evaporation from dust icy mantles during the outburst and its slower freeze-out afterwards. Apart from direct measurements of variations in gas-phase CO abundances, we may also hope to detect their indirect effect in the abundance of N$_2$H$^+$, which is efficiently destroyed by CO \citep{lee2007}, in the abundances of CO `daughter' species (like HCO$^+$ or H$_2$CO), or in deuterium isotopic chemistry. The other straightforward chemical consequence is the change in the CO and CO$_2$ ice abundances during and after the outburst, provided that effective conversion of CO to CO$_2$ occurs on dust grain surfaces during quiescent phases \citep{2011ApJ...729...84K,2012ApJ...758...38K}. More specifically, CO molecules, having smaller desorption energy, should evaporate during the outburst, leaving behind ice mantles, mostly consisting of less volatile CO$_2$. Such an excess of CO$_2$ ice has indeed been found in spectra of about half of low luminosity young stellar objects by \cite{2012ApJ...758...38K}.

Details of preferential CO desorption during the outburst in a protostellar object consisting of a dynamically evolving disc and envelope were considered by \cite{Vorobyov2013}. These authors have found that during a typical outburst CO$_2$ evaporates only in the inner part of the disc, while CO evaporation encompasses the disc and the most part of the envelope. A significant amount of CO stays in the gas-phase long after the outburst has faded. A more robust consideration of this process based on a detailed chemical model, but using a steady state disc and envelope model, was presented by \cite{2017A&A...604A..15R}. The authors demonstrated that post-outburst objects can be efficiently identified by an unusually large extent of their CO emission in the envelope.

Such an extended CO emission was detected in some low-mass protostars by \cite{2013ApJ...779L..22J,2015A&A...579A..23J} and \citet{Frimann2017}. In about half of their objects, the extent of CO emission exceeds limits which would be appropriate for current luminosities of these objects. The authors concluded that in the past $10^4$ years these objects experienced a transient rise in luminosity by at least a factor of five. Interestingly,  \cite{2013ApJ...779L..22J} found that HCO$^+$ emission in IRAS 15398-3359, unlike CO emission, is not seen in the vicinity of the protostar, rather forming a ring-like structure with a radius of about 150--200 au. They suggested that HCO$^+$ is destroyed by water molecules released from dust surfaces during a recent luminosity outburst.

This indicates that chemistry during and after an outburst is not limited to a simple cycle of evaporation and freeze-out. A detailed theoretical study of the outburst chemistry with a single-point model was performed by \cite{Visser2012}. They considered an effect of recurrent outbursts on the abundances of simple species, like CO, CO$_2$, HCO$^+$, and N$_2$H$^+$, and concluded that the gas-phase CO and N$_2$H$^+$, along with CO$_2$ ice, may serve as probes of episodic accretion in low-mass protostars. This study was followed by the work of \cite{Visser2015}, who considered a grid of models representing spherical protostellar envelopes with pre-set density and temperature profiles. Several line ratios for the same molecules were suggested as chronometers for episodic accretion.

Common features of the cited studies are that 1) the chemical effect of luminosity outbursts is considered for a limited range of densities usually corresponding to the outer accretion disc and protostellar envelope; 2) surface processes are either totally neglected or taken into account in a simplified manner. For example, the highest density that is considered in \cite{Visser2015} is only $2\times10^9$ cm$^{-3}$, while much higher densities are expected in the inner regions of protoplanetary discs. As a result, some effects that are related to dense material close to a protostar and to surface chemistry can be missed.

In this study, following \cite{Visser2012}, we use a single-point model to investigate the influence of a strong luminosity outburst onto the chemical composition of a protoplanetary disc within 3--250 au from the central source. A detailed chemical model is used, which takes into account surface reactions along with adsorption and desorption processes. We explore how molecular abundances respond to both heating and illumination related to the outburst and also reveal species (including organic ones), which are most sensitive to the outburst-related chemistry.

The paper is organized as follows. In Sect. 2 we provide a description for chemical and physical models used in this study and discuss initial conditions for the modelling. In Sect. 3 we describe results for the fiducial model that is intended to represent a ``typical'' disc with a ``typical'' luminosity outburst. Sect. 4 is devoted to the influence of possible deviations from the fiducial model. Our results are discussed in Sect. 5 and summarised in Sect. 6.

\section{Model description}

\subsection{Chemical model}

We utilise a single-point astrochemical model, based on the ALCHEMIC reaction network \citep{ALCHEMIC}. Both gas-phase and surface processes are considered in the model, along with the processes of adsorption onto dust grains and desorption from icy mantles. Included desorption processes are thermal desorption, photodesorption, and cosmic ray induced desorption. The UV photodesorption yield of $10^{-5}$ for all the species is adopted \citep{Cruz_Diaz2016,Bertin2016}. Cosmic ray particles, X-ray photons, and decay products of short-lived radionuclides are assumed to evaporate icy mantles, impulsively heating dust grains \citep{1985A&A...144..147L}. When modelling surface chemistry, we adopt the ratio of diffusion energy to desorption energy of 0.5 and account for tunnelling through reaction barriers. Stochastic effects are not accounted for as we do not expect them to be important given the high densities considered. Because the behaviour of benzene turns out to be interesting in some situations, several reactions from the latest version of the KIDA database \citep{KIDA} involving benzene have been added to the utilized network. Also, several reactions involving propynal (H$_2$C$_3$O) have been added from \cite{2016MNRAS.456.4101L}.

While astrochemical models, which take into account the presence of dust grains having various sizes, are available \citep[see e.g.][]{2011ApJ...732...73A,2014ARep...58..228K,2016ApJ...817..146P}, in the adopted formalism for surface reactions \citep{hhl92} only total grain surface area per unit volume matters \citep{2011ApJ...732...73A}. In this study, dust grains are assumed to have a power law size distribution \citep[MRN,][]{mrn} with a lower limit of 0.005 $\mu$m. An upper limit in the fiducial model is taken to be 1~$\mu$m, which corresponds to a `typical' grain size of about $10^{-6}$ cm \citep{2017A&A...604A..15R}. In some models, we vary the upper limit in order to explore the effects of grain evolution in protoplanetary discs.

\subsection{Physical conditions}

A wide range of conditions is expected to occur in a typical protoplanetary disc, but in order to capture main trends in the chemical evolution prior, during, and after an outburst, we only need to consider a limited set of typical combinations of gas density, $n$, gas and dust temperature, $T$, ionization rate related to energetic particles, $\zeta$, and the normalized radiation field, $\chi$. Considered combinations of physical parameters are summarised in Table~\ref{tab:physcon}. We assume that gas and dust temperatures are equal. This assumption is correct in the dense disc regions but almost certainly fails in the disc atmosphere. However, we believe that our results are not compromised by this assumption as dust and gas temperatures start to diverge only in the low-density transparent disc regions, where chemistry is dominated by photoprocesses.

\begin{table*}
	\centering
	\caption{Physical conditions in the representative disc locations.}
	\label{tab:physcon}
\begin{tabular}{l|llll|llll}
\hline
           & \multicolumn{4}{c}{Dark Disc}& \multicolumn{4}{c}{Atmosphere}\\
\hline
Disc region, phase & Density, cm$^{-3}$ & $T$, K & $\chi$& $\zeta$, s$^{-1}$ & Density, cm$^{-3}$ & $T$, K & $\chi$& $\zeta$, s$^{-1}$ \\
\hline
Inner, quiescent  & $8\times10^{11}$ & 100 & 0 & $10^{-15}$ & $1.5\times10^{10}$ & 160 & 0 & $10^{-12}$\\
Inner, outburst  & $8\times10^{11}$ & 500 & 0 & $10^{-15}$ & $1.5\times10^{10}$ & 800 & $10^4$ & $10^{-12}$\\
\hline
Intermediate, quiescent  & $2\times10^{11}$ & 50  & 0 & $10^{-16}$ & $2\times10^{9}$ & 80 & 0 & $10^{-13}$\\
Intermediate, outburst  & $2\times10^{11}$ & 200 & 0 & $10^{-16}$& $2\times10^{9}$ & 300 & $10^3$ & $10^{-13}$\\
\hline
Outer, quiescent & $4\times10^{9}$ & 20  & 0& $10^{-17}$ & $6\times10^{7}$ & 40 & 0 & $10^{-17}$\\
Outer, outburst  & $4\times10^{9}$ & 80 & 0 & $10^{-17}$& $6\times10^{7}$ & 100 & 150 & $10^{-17}$\\
\hline
Extreme outer, quiescent & $2\times10^{7}$ & 10  & 0 & $10^{-17}$& $5\times10^{5}$ & 30 & 0 & $10^{-17}$\\
Extreme outer, outburst  & $2\times10^{7}$ & 40 & 0 & $10^{-17}$& $5\times10^{5}$ & 50 & 100 & $10^{-17}$\\
\hline
	\end{tabular}
\end{table*}

As a guide for physical conditions, we use a parametric disc model, described in \cite{2017ApJ...849..130M}. A disc has a pre-set surface density radial profile and is illuminated by the radiation field of the central star, whose parameters were derived from \cite{2015A&A...577A..42B} models, accretion luminosity corresponding to a specific value of $\dot{M}$, and interstellar radiation field \citep{1983A&A...128..212M}. In our fiducial model, a quiescent state of the disc is approximately represented by $\dot{M}=10^{-9}\,M_{\sun}$ yr$^{-1}$, while an outburst corresponds to $\dot{M}=10^{-5}\,M_{\sun}$ yr$^{-1}$.

\begin{figure}
\includegraphics[width=\columnwidth]{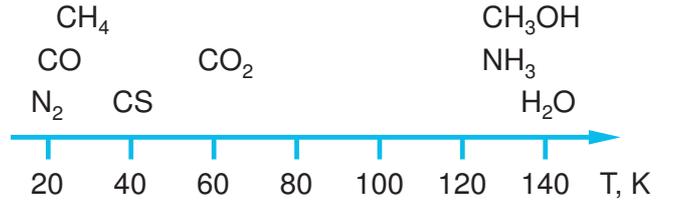}
    \caption{Scheme showing sublimation temperatures for some species considered in the paper.}
    \label{destemp}
\end{figure}

We use the disc model to select representative combinations of $n$, $T$, $\zeta$, and $\chi$. For the `dark disc' conditions, roughly corresponding to the disc midplane, we consider four locations, having temperatures of 100\,K, 50\,K, 20\,K, and 10\,K. In the adopted disc model they roughly correspond to radial distances of a few au, $\approx10$ au, $\approx100$ au, and $\approx250$ au. In the following, these locations are designated as an inner dark disc region, an intermediate dark disc region, an outer dark disc region, and an extreme outer dark disc region. The temperature in the inner disc region is higher than the sublimation temperature for most species, so that only water, ammonia, and some complex organic molecules are abundant on dust surfaces at a quiescent state, excluding those components (like methanol) that form on dust surfaces out of lighter compounds. The intermediate region temperature (50\,K, 10~au) is higher than the CO sublimation temperature, but lower than the CO$_2$ sublimation temperature \citep{2015A&A...582A..41H}. It corresponds to the borderline situation when carbon monoxide still mostly resides in the gas phase, but can be gradually converted to CO$_2$ on dust surfaces, thus locking some carbon atoms in the solid phase. The temperature in the outer dark disc is lower than sublimation temperatures for most species, allowing for the formation of chemically diverse icy mantles. Finally, the extreme outer dark disc temperature roughly corresponds to the edge of the disc. On Fig.~\ref{destemp} we show some important species with sublimation temperatures roughly corresponding to the considered disc regions. These sublimation temperatures are computed for the adopted desorption energies listed in Table \ref{desen}.

In addition, we also consider `atmospheric' locations at the same four radii roughly at the height that is mostly dark in the quiescent state and is assumed to be illuminated during the outburst. This assumption may actually be somewhat unrealistic as a disc (at least, in the framework of the used model) puffs up in the response to the luminosity outburst, so that parcels of matter that have been dark before the outburst stay dark during the outburst. But we neglect this vertical expansion in order to check whether an increased luminosity may cause some observable effects. A more profound influence of the outburst onto the radiation field is expected at the uppermost disc layers, but they are already well illuminated at the quiescent state, so that the increase in the luminosity is not that important. For uniformity, we assume zero illumination during the quiescent and post-outburst stages everywhere in the disc, even though interstellar UV penetration can be a significant factor in the outermost disc regions. On the other hand, at this early evolutionary stage remnants of a pre-stellar envelope may prevent interstellar UV photons from getting into the disc.

We assume that within 10 au from the star the ionization rate is mostly due to stellar X-ray photons, while in the outer and extreme outer disc regions the `canonical' cosmic ray ionization rate of $10^{-17}$~s$^{-1}$ is used. The energetic cosmic rays and X-rays equally interact with matter both directly, through ionization, and indirectly, through UV field induced by H$_2$ excitation by secondary electrons \citep{1987ApJ...323L.137G}. For brevity in the following we shall collectively refer to this UV field as XCR-induced photons, keeping in mind that in the model they are in fact induced both by cosmic rays and by X-rays.

\subsection{Initial conditions}

We start the computation from a molecular cloud stage, which lasts for $10^6$ years and is needed to establish an initial inventory of gas-phase and solid-phase molecules. The chemical evolution at this stage is computed for a gas density of $10^4$~cm$^{-3}$ and temperature of 10\,K. At the beginning, gas is assumed to have atomic initial composition with the only exception of hydrogen, which is taken to be almost entirely molecular. For heavier atoms, we use the so-called low-metal abundance set from the work of \cite{1998A&A...334.1047L}. All the initial abundances are listed in Table~\ref{inichem}. By the end of the molecular cloud stage nearly all oxygen, carbon, and nitrogen atoms are locked in surface water, methanol, and ammonia, respectively (see Table~\ref{molchem}). Most abundant gas-phase molecules are CO, O$_2$, and N$_2$ with relative abundances of the order of $10^{-6}$.

\begin{table}
\caption{Initial elemental abundances relative to the number of H nuclei. Notation $A(B)$ means $A\times10^{B}$.}
\label{inichem}
\begin{tabular}{ll}
\hline
Species & Abundance \\
\hline
H            &       1(--5)       \\
H$_2$        &       4.99995(--1) \\
C            &       7.30(--5)    \\
N            &       2.14(--5)    \\
O            &       1.76(--4)    \\
S            &       8.00(--8)    \\
Si           &       8.00(--9)    \\
Fe           &       3.00(--9)    \\
Na           &       2.00(--9)    \\
Mg           &       7.00(--9)    \\
Cl           &       1.00(--9)    \\
P            &       2.00(--10)   \\
\hline
\end{tabular}
\end{table}

\begin{table}
\caption{Abundances of selected species relative to the number of H nuclei after the molecular cloud stage. Notation $A(B)$ means $A\times10^{B}$.}
\label{molchem}
\begin{tabular}{llll}
\hline
Species & Abundance & Species & Abundance \\
\hline
s-H$_2$O      & 1.2(--4) &  CH$_3$        & 4.6(--8) \\ 
H             & 5.8(--5) &  s-C$_4$H$_4$  & 4.3(--8) \\
s-CH$_3$OH    & 3.9(--5) &  H$_2$O$_2$    & 3.6(--8) \\
s-CH$_4$      & 1.8(--5) &  s-CH$_3$CN    & 2.7(--8) \\
s-NH$_3$      & 1.3(--5) &  H$_2$CO       & 2.3(--8) \\
CO            & 8.8(--6) &  NH$_2$        & 1.9(--8) \\
s-HCN         & 2.2(--6) &  CH$_2$OH      & 1.8(--8) \\
O$_2$         & 2.1(--6) &  O$_3$         & 1.7(--8) \\
N$_2$         & 1.8(--6) &  C             & 1.6(--8) \\
O             & 1.1(--6) &  s-SO          & 1.6(--8) \\
s-C$_2$H$_6$  & 7.5(--7) &  s-CO          & 1.6(--8) \\
s-N$_2$       & 5.1(--7) &  HNO           & 1.4(--8) \\
s-C$_3$H$_4$  & 4.5(--7) &  CH$_3$OH      & 1.4(--8) \\
CH$_4$        & 4.3(--7) &  C2H$_2$       & 1.2(--8) \\
N             & 3.7(--7) &  s-C$_5$H$_4$  & 1.1(--8) \\
NO            & 2.9(--7) &  H$_3^+$       & 1.1(--8) \\
OH            & 2.6(--7) &  SO            & 8.7(--9) \\
H$_2$O        & 2.2(--7) &  CN            & 8.0(--9) \\
s-HNCO        & 1.9(--7) &  s-MgH$_2$     & 6.5(--9) \\
s-CO$_2$      & 1.6(--7) &  HCN           & 6.3(--9) \\
s-HNC         & 1.5(--7) &  s-NH$_2$CHO   & 5.7(--9) \\
s-CH$_3$NH$_2$& 8.8(--8) &  HNC           & 5.2(--9) \\
s-H$_2$C$_3$O & 7.6(--8) &  HC$_3$O       & 5.1(--9) \\
s-C$_2$H$_5$CN& 6.9(--8) &  s-SiH$_4$     & 4.9(--9) \\
OCN           & 6.6(--8) &  H$_3$O$^+$    & 4.9(--9) \\
NH$_3$        & 6.4(--8) &  s-C$_6$H$_4$  & 3.6(--9) \\
s-H$_2$CO     & 5.8(--8) &  NH            & 2.7(--9) \\
CO$_2$        & 5.0(--8) &  s-FeH         & 2.3(--9) \\
s-CH$_2$CO    & 4.9(--8) &  H$_2$S        & 2.2(--9) \\
S             & 4.7(--8) &  HCO$^+$       & 2.2(--9) \\
\hline                                 
\end{tabular}
\end{table}

The methanol ice being the dominant C-bearing compound after the end of the molecular stage is in apparent contradiction with observational data on ice abundances in protostellar objects. The content of methanol ice relative to water ice exceeds 30\% (Table~\ref{molchem}) in our model, while the relative content of CO$_2$ ice is only 0.13\%, and the CO ice abundance is even smaller, while observationally inferred abundances for CO and CO$_2$ ices are of the order of 10\%--30\% relative to water ice \citep[see, e.g.][]{2011ApJ...740..109O,2018A&A...610A...9G}. This is because in our model surface CO is effectively converted to formaldehyde and then to methanol. In order to suppress somewhat the surface methanol synthesis we have added the backward hydrogen abstraction reaction s-H~+~s-CH$_3$OH to the network with the parameters taken from \citet{2013ApJ...765...60G}, but this does not alter the molecular stage abundances appreciably due to high activation energy for this reaction. The issue is further discussed in Section 4.4.

The chemical composition from the molecular cloud stage is used as the initial condition for a pre-outburst quiescent disc stage, which lasts for another $5\times10^5$ years in the fiducial model (the influence of the quiescent stage duration is considered in Section 4). At this stage, the initial composition adapts to higher density and higher temperature. 

In the inner dark disc the quiescent pre-outburst evolution at a high density ($8\cdot10^{11}$~cm$^{-3}$) and temperature ($T=100$~K) alters the initial molecular abundances most significantly. Nearly all carbon atoms reside in the gas-phase CO and CO$_2$ molecules, while O atoms are shared between surface water and gas-phase O$_2$, CO and CO$_2$ molecules. Grain mantles consist almost exclusively of water ice with a few per cent admixture of ammonia and some small hydrocarbons. Complex surface organic molecules by the end of the quiescent stage are mostly destroyed by photoreactions with the XCR-induced photons.

The intermediate dark disc, due to its lower temperature (50~K), is characterised by a richer mantle composition by the end of the quiescent stage. Water is still the most abundant ice component, but farther from the star it is mixed with more appreciable amounts of ammonia, CO$_2$, and ethane ices. The latter three molecules together account for about 30\% of surface species (by number). More complex organics, like methylamine (CH$_3$NH$_2$), methanol (CH$_3$OH), formic acid (HCOOH), and acetaldehyde (CH$_3$CHO) are also present, but at a low level, being mostly destroyed by the XCR-induced photons. The most abundant gas-phase species (apart from H$_2$ and helium) are atomic hydrogen, CO, and methane.

In the cold and relatively dense medium of the outer and extreme outer disc, nearly all molecules are frozen out during the pre-outburst stage. The icy mantles consist mostly of water, methane, ammonia, methanol, carbon dioxide, and formaldehyde ices. At the disc periphery, the adopted CR ionization rate is set to its canonical Galactic value, without any X-ray contribution. Thus, the surface XCR-induced processes are less effective here than closer to the star, and complex organic molecules are also abundant in the grain mantles. The most abundant gas-phase compound (apart from H$_2$ and He) is molecular nitrogen with an abundance of about $3\div4\times10^{-9}$.

In the fiducial model, the pre-outburst stage is followed by a luminosity outburst with a duration of 100 years. The outburst is assumed to start and stop instantaneously. After the outburst a post-outburst quiescent stage commences. We follow the post-outburst evolution for another $10^4$ years, using the same physical conditions as at the pre-outburst stage. Only molecules with abundances greater than $10^{-12}$ during at least one of the stages are considered.

\section{Results for the fiducial model}

In this section, we present results for a fiducial model, which we consider as a reference point to study chemical effects of a luminosity outburst. Then, in the next section, we proceed with presenting results for models, in which we vary some parameters within limits consistent with observational, experimental and/or theoretical constraints. Considered are abundances at the end of the quiescent stage ($x_{\rm q}$), at the end of the outburst ($x_{\rm o}$), and 500 years after the outburst ($x_{\rm p}$).

All the species can be divided into five categories according to their behaviour during and after the outburst. A {\em zero type} response implies no noticeable response at all. A {\em first type\/} encompasses species, whose abundances change sharply at the beginning of the outburst, do not change appreciably during the outburst, and return rapidly to pre-outburst values after the end of the outburst. Mathematically, the following conditions apply: $R_{\rm pq} \in [0.1,10]$, $R_{\rm oq} \notin [0.1,10]$, $R_{\rm oo} \in [0.1,10]$, where $R_{\rm pq}=x_{\rm p}/x_{\rm q}$, $R_{\rm oq}=x_{\rm o}/x_{\rm q}$, and $R_{\rm oo}$ is the ratio of abundances calculated 5 yr into the outburst and at its end. In other words, the abundance evolution for these species is described by a $\Pi$-like profile (or an upturned $\Pi$-like profile). These species can serve as immediate tracers of the outburst but do not retain any memory of the event after it has completed. As a rule, their abundances during the outburst are controlled by some unique process like mantle evaporation or photo-ionization.

A {\em second type\/} of behaviour is related to cases when abundances not only change rapidly at the very start of the outburst, but also evolve significantly during the whole period of increased luminosity ($R_{\rm pq} \in [0.1,10]$, $R_{\rm oq} \notin [0.1,10]$, $R_{\rm oo} \notin [0.1,10]$). This implies that the raised temperature (and UV irradiation) initiate reaction sequences that are more complicated than those related to the first type. Still, after the end of the outburst the abundances of these species quickly return to their pre-outburst values. They also can only be used as tracers of an ongoing outburst, while being important as precursors for processes of the third type.

Post-outburst abundances of species experiencing response of a {\em third type\/} differ significantly from their pre-outburst abundances and stay at this altered level for centuries after the end of the outburst. These abundances either preserve the high values attained during the outburst or even grow after the outburst has subsided ($R_{\rm pq} \geq 10$). Thus, species showing the third type of behaviour are the most promising tracers of the past outburst activity.

For the sake of completeness, we also define a {fourth type\/} for species irreversibly destroyed by the outburst ($R_{\rm pq} \leq 0.1$). To reveal the past outburst in the ideal case, one would want to observe species of the third type, while seeing no sign of species of the fourth type.

Two additional remarks should be made. First, same species may demonstrate different types of behaviour at different locations. Second, there are no species in our models (except for He) that would not have responded to the outburst in some way or another at least in one of the considered locations (we assume that there is a response if the abundance of the molecule changes by more than an order of magnitude).j u

\subsection{Dark disc}

\subsubsection{First type}

\begin{table}
\caption{Abundances of species of the first type in the dark disc, sorted by the decreasing $x_{\rm o}$ value. In the last column a location code is indicated: I --- inner disc, IM --- intermediate disc, O --- outer disc.}
\label{ddgr1}
\begin{tabular}{llllll}
\hline
Species    & $x_{\rm q}$   & $x_{\rm o}$ &  $x_{\rm p}$ & $R_{\rm oq}$ & Loca-\\
 &   &  &   & & tion\\
\hline
H$_2$O      & 8.8(--14) & 1.4(--04) & 8.9(--14) & 1.5(+09) & IM \\
H$_2$O      & 4.1(--12) & 1.1(--04) & 2.9(--12) & 2.7(+07) & I \\
CH$_4$      & 3.4(--12) & 2.8(--05) & 3.4(--12) & 8.1(+06) & O \\
NH$_3$      & 2.9(--14) & 2.1(--05) & 3.3(--14) & 7.3(+08) & IM \\
CO$_2$      & 4.1(--10) & 1.9(--05) & 4.1(--10) & 4.8(+04) & IM \\
C$_2$H$_6$  & 3.9(--14) & 1.5(--05) & 3.9(--14) & 3.9(+08) & IM \\
CO$_2$      & 1.3(--13) & 1.1(--05) & 3.0(--13) & 8.3(+07) & O \\
H$_2$CO     & 1.8(--15) & 1.1(--05) & 1.9(--15) & 6.0(+09) & O \\
C$_3$H$_4$  & 5.7(--14) & 5.5(--06) & 5.2(--14) & 9.6(+07) & IM \\
NH$_3$      & 2.2(--13) & 2.8(--06) & 2.2(--13) & 1.3(+07) & I \\
HCN         & 1.6(--14) & 2.4(--06) & 1.6(--14) & 1.5(+08) & O \\
N$_2$       & 4.6(--09) & 2.2(--06) & 4.6(--09) & 4.9(+02) & O \\
C$_5$H$_4$  & 6.2(--15) & 8.7(--07) & 6.2(--15) & 1.4(+08) & I \\
HNO         & 6.0(--13) & 3.4(--07) & 6.1(--13) & 5.7(+05) & O \\
C$_3$H$_2$  & 8.3(--15) & 2.9(--07) & 1.3(--14) & 3.6(+07) & IM \\
C$_4$H$_4$  & 2.7(--12) & 1.2(--07) & 1.0(--12) & 4.5(+04) & I \\
C$_4$H$_4$  & 7.0(--16) & 1.1(--07) & 1.1(--15) & 1.5(+08) & IM \\
HNCO        & 7.8(--21) & 9.8(--08) & 8.2(--21) & 1.3(+13) & O \\
NH$_2$OH    & 1.5(--16) & 9.6(--08) & 2.1(--16) & 6.3(+08) & I \\
H$_2$S      & 1.7(--12) & 7.6(--08) & 1.8(--12) & 4.6(+04) & O \\
H$_2$CS     & 1.5(--13) & 7.4(--08) & 1.5(--13) & 4.8(+05) & IM \\
C$_6$H$_4$  & 4.3(--16) & 7.0(--08) & 1.0(--15) & 1.6(+08) & I \\
SO$_2$      & 8.1(--12) & 3.8(--08) & 5.9(--12) & 4.7(+03) & I \\
CH$_2$CO    & 4.0(--20) & 3.5(--08) & 3.9(--20) & 8.6(+11) & O \\
CO          & 6.4(--15) & 2.2(--08) & 2.2(--14) & 3.4(+06) & O \\
H           & 9.3(--11) & 2.1(--08) & 9.3(--11) & 2.2(+02) & O \\
C$_7$H$_4$  & 1.3(--16) & 2.0(--08) & 1.6(--16) & 1.5(+08) & I \\
CH$_3$NH$_2$& 8.6(--18) & 1.9(--08) & 9.4(--18) & 2.2(+09) & IM \\
NH$_2$      & 1.4(--14) & 1.6(--08) & 2.4(--14) & 1.1(+06) & IM \\
C$_3$H$_3$  & 6.6(--19) & 1.5(--08) & 6.2(--18) & 2.2(+10) & I \\
CH$_3$CHO   & 9.0(--16) & 1.3(--08) & 8.9(--16) & 1.5(+07) & IM \\
C$_5$H$_4$  & 9.5(--17) & 1.2(--08) & 3.4(--16) & 1.3(+08) & IM \\
HCOOH       & 9.0(--16) & 9.6(--09) & 3.7(--16) & 1.1(+07) & I \\
H$_2$C$_3$O & 8.3(--17) & 8.5(--09) & 1.1(--16) & 1.0(+08) & IM \\
HCOOH       & 1.7(--18) & 8.5(--09) & 1.9(--18) & 5.0(+09) & IM \\
H$_2$C$_3$O & 6.2(--24) & 8.2(--09) & 7.5(--24) & 1.3(+15) & O \\
MgH$_2$     & 1.4(--17) & 7.0(--09) & 1.4(--17) & 5.0(+08) & IM \\
MgH$_2$     & 1.0(--17) & 6.9(--09) & 1.0(--17) & 6.7(+08) & I \\
SiH$_4$     & 7.7(--18) & 6.6(--09) & 7.7(--18) & 8.5(+08) & IM \\
CH$_3$OH    & 2.5(--17) & 5.6(--09) & 2.7(--17) & 2.2(+08) & IM \\
\hline
\end{tabular}
\end{table}

\begin{figure}
\includegraphics[width=\columnwidth]{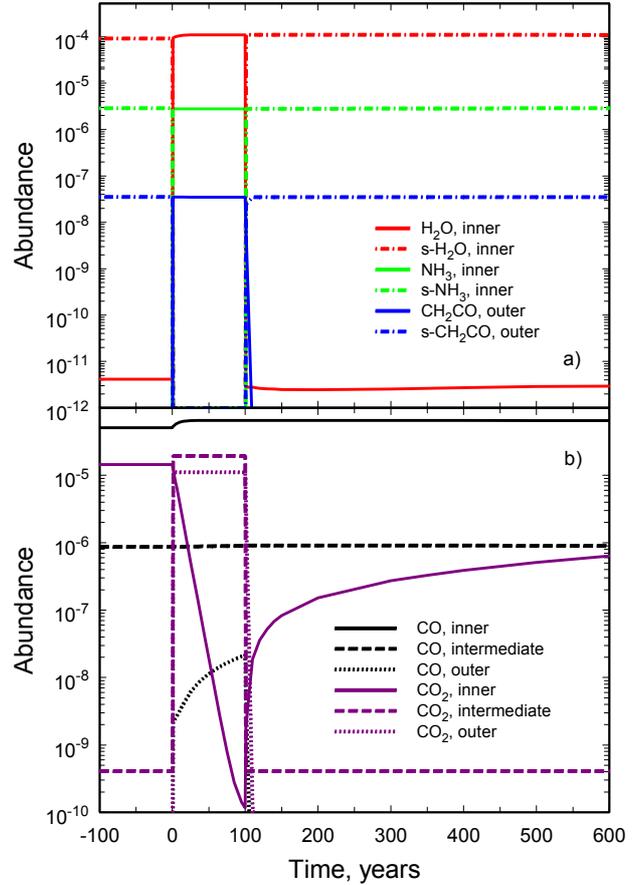}
    \caption{Evolution of some species of the first type in various dark disc regions through the outburst. The outburst starts at $t=0$ years and lasts for 100 years. Surface species are indicated with `s-'.}
    \label{wiebe_fig02}
\end{figure}

In the dark disc the sole action of the outburst on species of the first type is their rapid evaporation from icy mantles followed by the equally rapid freeze-out after the end of the outburst. This effect of the outburst is especially straightforward in the inner and intermediate disc regions, where the rise in the dust temperature causes the evaporation of even less volatile species, leading to a rapid growth of their gas-phase abundances. In Table~\ref{ddgr1} we list species of the first type with the highest abundances during the outburst, and examples of their abundance variations are shown in Fig.~\ref{wiebe_fig02}a. Note that the total (gas+surface) abundance of each species stays nearly constant. The only outburst-driven change is the evaporation-readsorption cycle.

It is interesting that neither CO, nor CO$_2$, which are often considered as outburst markers, do not actually serve this purpose in the vicinity of the star (Fig.~\ref{wiebe_fig02}b). Nothing happens to CO in the inner and intermediate disc regions during the outburst as dust temperature is so high there even at the quiescent stage that all the CO is in the gas-phase anyway. Evaporation of CO produce a noticeable effect only in the outer disc, but its gas-phase abundance during the outburst is quite modest there, making it hard to detect. Some growth of the CO abundance during the outburst is in part caused by reactions with atomic hydrogen (e.g. H~+~HCO), as H abundance also grows, but this increase does not exceed an order of magnitude, so, according to our definition, CO still falls into the first type.

Technically speaking, CO$_2$ does not belong to the first type in the inner disc. Its abundance {\em decreases\/} during the outburst due to collisions with H$_2$ molecules, dropping from $2\times10^{-5}$ in the beginning of the outburst to less than $10^{-9}$ by its end (Fig.~\ref{wiebe_fig02}b). While a barrier of this reaction is quite high (7550\,K in the adopted network), it is effective due to high density. We mostly mention this reaction as an indication of some possible unexpected offshots of high-temperature chemistry in the inner regions of outbursting protostars. It is considered in some AGB star models (N. Harada, private communication), but is missing in publicly available astrochemical ratefiles. CO$_2$ destruction in the inner disc is the only effect of this reaction in the presented model. After the end of the outburst abundance of CO$_2$ rapidly grows, but still is less than 10\% of its pre-outburst value (within 500 years). There are also other gas-phase species that are destroyed by high-T chemistry during the outburst in the inner disc, mostly in reactions with atomic H. These are O$_2$, SiO, N$_2$O, and SiO$_2$. Together with CO$_2$, they rather comprise the fourth type in the dark disc.

Farther out from the star, CO$_2$ behaves as a typical molecule of the first type. The period of enhanced luminosity is marked by a sharp increase in its abundance, which, unlike in the inner disc, is not decreased here by the collisions with H$_2$ due to less effective outburst heating and smaller density. Neither CO, nor CO$_2$ show any memory of the outburst everywhere except for the extreme outer disc as discussed below in Section 3.1.3. Note that no abundant molecules show the behaviour of the first type in the extreme outer disc as such a rapid evolution is possible mostly in the vicinity of the star.

\subsubsection{Second type}

\begin{table}
\caption{Abundances of species of the second type in the dark disc, sorted by the decreasing $x_{\rm o}$ value. In the last column a location code is indicated: I --- inner disc, IM --- intermediate disc, O --- outer disc, EO --- extreme outer disc.}
\label{ddgr2}
\begin{tabular}{llllll}
\hline
Species    & $x_{\rm q}$   & $x_{\rm o}$ &  $x_{\rm p}$ & $R_{\rm oq}$ & Loca-\\
 &   &  &   & & tion\\
\hline
C$_2$H$_2$   & 6.2(--11) & 1.0(--06) & 3.9(--10) & 1.7(+04) & I  \\
C$_6$H$_6$   & 4.2(--20) & 5.0(--07) & 9.2(--20) & 1.2(+13) & IM \\
C$_6$H$_2$   & 6.9(--20) & 7.1(--08) & 3.4(--19) & 1.0(+12) & IM \\
C$_2$H$_2$   & 2.8(--14) & 6.1(--08) & 4.0(--14) & 2.2(+06) & IM \\
C$_4$H$_2$   & 1.0(--15) & 4.6(--08) & 1.4(--15) & 4.6(+07) & IM \\
CH$_3$       & 1.4(--11) & 3.3(--08) & 1.7(--11) & 2.3(+03) & IM \\
C$_5$H$_2$   & 4.8(--18) & 3.2(--08) & 1.1(--17) & 6.7(+09) & IM \\
C$_2$H$_4$   & 6.9(--14) & 3.1(--08) & 1.0(--13) & 4.5(+05) & IM \\
NO           & 5.9(--13) & 1.9(--08) & 6.1(--13) & 3.2(+04) & O  \\
HC$_3$N      & 6.6(--18) & 7.3(--09) & 3.2(--15) & 1.1(+09) & I  \\
HNC          & 2.7(--16) & 6.6(--09) & 2.7(--16) & 2.4(+07) & O  \\
HC$_3$O      & 1.0(--16) & 4.8(--09) & 1.6(--16) & 4.7(+07) & IM \\
C$_7$H$_2$   & 1.2(--21) & 3.8(--09) & 4.4(--21) & 3.3(+12) & IM \\
NO           & 1.1(--10) & 3.8(--09) & 4.7(--10) & 3.5(+01) & EO \\
C$_3$O       & 4.0(--17) & 3.7(--09) & 5.2(--17) & 9.2(+07) & IM \\
C$_8$H$_2$   & 2.3(--22) & 3.1(--09) & 1.5(--21) & 1.3(+13) & IM \\
C$_6$H$_6$   & 1.6(--18) & 2.3(--09) & 7.6(--18) & 1.4(+09) & I  \\
s-C$_9$H$_2$ & 4.5(--15) & 1.9(--09) & 5.3(--14) & 4.2(+05) & IM \\
C$_9$H$_2$   & 8.6(--21) & 1.5(--09) & 1.1(--20) & 1.8(+11) & IM \\
NS           & 2.6(--17) & 1.5(--09) & 6.7(--15) & 5.6(+07) & I \\
HC$_5$N      & 1.1(--18) & 1.2(--09) & 3.8(--17) & 1.1(+09) & I \\
CH$_2$NH     & 7.6(--21) & 4.0(--10) & 4.0(--18) & 5.3(+10) & I \\
Na           & 5.1(--22) & 2.3(--10) & 6.2(--22) & 4.6(+11) & I  \\
CH$_3$C$_6$H & 7.4(--22) & 1.0(--10) & 2.6(--21) & 1.4(+11) & IM \\
CH$_3$NH$_2$ & 2.9(--21) & 1.0(--10) & 1.7(--19) & 3.5(+10) & I \\
s-C$_8$H$_2$ & 7.1(--15) & 7.5(--11) & 5.1(--14) & 1.1(+04) & IM \\
MgH          & 1.4(--24) & 6.6(--11) & 1.4(--24) & 4.6(+13) & I  \\
PN           & 5.7(--15) & 4.0(--11) & 4.2(--15) & 7.0(+03) & I  \\
CS           & 2.8(--14) & 3.8(--11) & 2.9(--14) & 1.4(+03) & IM \\
S            & 6.8(--15) & 1.3(--11) & 7.3(--15) & 1.9(+03) & O  \\
CH$_2$NH     & 2.4(--20) & 1.0(--11) & 2.8(--20) & 4.3(+08) & IM \\
\hline
\end{tabular}
\end{table}

\begin{figure}
\includegraphics[width=\columnwidth]{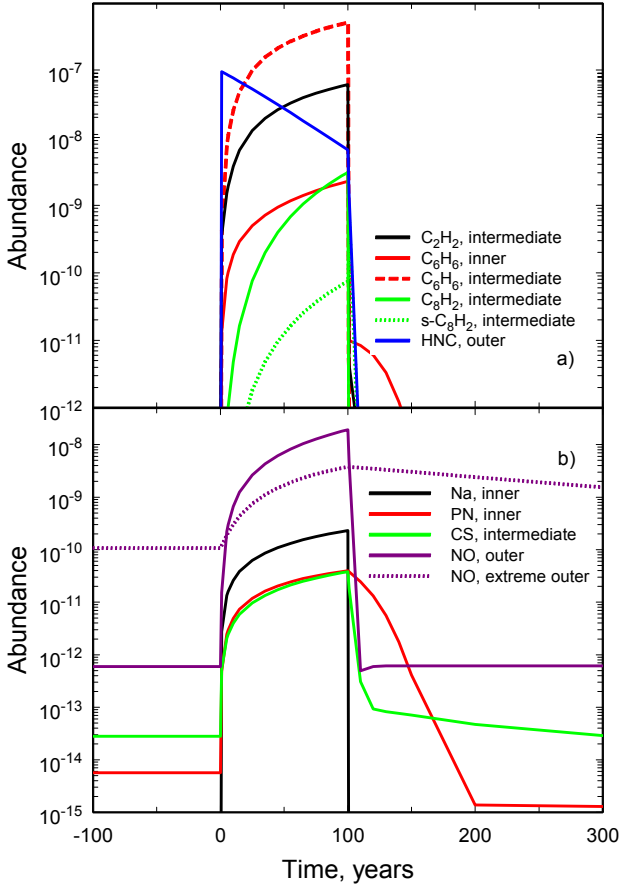}
    \caption{Evolution of some abundances of the species of the second type in various dark disc regions through the outburst. The outburst starts at $t=0$ years and lasts for 100 years. Surface species are indicated with `s-'.}
    \label{wiebe_fig03}
\end{figure}

In Table~\ref{ddgr2} we provide a list of species with the highest $x_{\rm o}$ values, related to the second type. Some species are also added to the list that formally belong to the third type, like HC$_3$N. However, their pre- and post-outburst abundances are extremely low so that even large difference between $x_{\rm o}$ and $x_{\rm p}$ does not bear any significance. Thus, efficiently these species belong to the second type. This list shows a striking prevalence of simple hydrocarbons in the inner and intermediate disc regions. At the quiescent stage, carbon atoms, initially locked in methanol ice, are released into the mantles due to surface methanol dissociation by the XCR-induced photons. Some of these carbon atoms combine into complex carbon-bearing species, up to s-C$_9$H$_4$. Most abundant carbon chain in the inner dark disc region before the outburst is s-C$_5$H$_4$. In the intermediate region, somewhat lower dust temperature suppress surface synthesis of s-C$_5$H$_4$ in favour of H addition reactions, leading to formation of abundant ethane ice (s-C$_2$H$_6$), cf. Table~\ref{ddgr1}. When the outburst commences, these species are evaporated and undergo some gas-phase processing. High-T reactions of C$_5$H$_4$ with H and H$_2$ in the inner region effectively convert it into acetylene (C$_2$H$_2$). Ion-molecular reactions initiated by the appearance of C$_5$H$_4$ and C$_2$H$_6$ in the gas-phase eventually lead to formation of C$_6$H$_7^+$, which dissociatively recombines into stable benzene molecules (Fig.~\ref{wiebe_fig03}a).

Surface chemistry of hydrocarbons also intensifies during the outburst in the vicinity of the star. This is why abundances of longer carbon chains also grow. We show C$_8$H$_2$ as an example in Fig.~\ref{wiebe_fig03}a. Both surface and gas-phase abundances of this species increase during the outburst in the intermediate disc region. Its major formation route is represented by surface carbon addition reactions that are followed by reactive desorption and thermal desorption.

Also shown in Fig.~\ref{wiebe_fig03}a is HNC that represents an example of a molecule, whose gas-phase abundance first grows and then decreases during the outburst. At the quiescent phase, HNC mostly resides on dust surfaces in the outer disc region, and its abundance is governed by the balance between the surface formation reaction s-C~+~s-NH and the destruction by XCR-induced photons. During the outburst this balance is shifted toward destruction due to decreased surface abundance of volatile s-C.

More examples of the outburst-driven evolution are presented in Fig.~\ref{wiebe_fig03}b. Sodium atoms appear in the gas-phase due to dissociation of NaH by XCR-induced photons. Sodium hydride is synthesized on dust surfaces, and this reaction is suppressed during the outburst in the inner disc region due to high gas temperature. In a similar way, CS is a product of H$_2$CS XCR-induced photodissociation, and the major (re)formation route of H$_2$CS is the surface s-H~+~s-HCS reaction that slows down during the outburst. PN is produced in a reaction between HPN$^+$ and NH$_3$, and this reaction intensifies during the outburst because of ammonia ice evaporation (cf. Fig.~\ref{wiebe_fig02}a). One should keep in mind that the phosphorus chemistry is quite limited in the utilized network, so that our results on P-bearing molecules should be treated with caution. The role of NH$_3$ in stabilizing big molecular ions by accepting a proton was also presented in \citet{2016ApJ...821...46T}.

The evolution of NO is also controlled by the balance between the synthesis and destruction of its parent molecule, HNO, which is shifted toward destruction (and, correspondingly, toward NO formation) during the outburst. The marked difference is that after the outburst the quiescent NO abundance in the extreme outer disc is restored much slower than in the case of other molecules. In 500 yr the NO abundance exceeds its pre-outburst abundance by less than an order of magnitude, which formally relates it to the second type. But actually abundances of most other molecules of this type are restored after the end of the outburst much faster. In this respect, the NO behaviour makes it similar to the molecules of the third type.

\subsubsection{Third type}

Species of the first and second types fade away rapidly after the outburst ends. This limits their utility as outburst tracers, since there are other signatures of current outbursts, such as the luminosity itself. From the observational point of view, more interesting are species that retain memory of the outburst for centuries and may allow detecting an outbursting star long after the outburst has ended.

In Table~\ref{ddgr3} we present species that are still over-abundant relative to the quiescent phase even 500 years after the outburst. To limit the size of the table, we only include species with $x_{\rm p}>10^{-12}$ and $R_{\rm pq}>10^2$. Note that this type is mostly populated by species from the extreme outer disc, especially when it concerns to gas-phase components.

\begin{table}
\caption{Abundances of species of the third type in the dark disc, sorted by the decreasing $x_{\rm p}$ value. In the last column a location code is indicated: I --- inner disc, IM --- intermediate disc, O --- outer disc, EO --- extreme outer disc.}
\label{ddgr3}
\begin{tabular}{llllll}
\hline
Species    & $x_{\rm q}$   & $x_{\rm o}$ &  $x_{\rm p}$ & $R_{\rm pq}$ & Loca-\\
 &   &  &   & & tion\\
\hline
CH$_4$            & 8.8(--10) & 2.9(--05) & 9.5(--07) & 1.1(+03) & EO \\
s-C$_6$H$_6$      & 1.8(--13) & 3.3(--08) & 5.3(--07) & 2.9(+06) & IM \\
CO                & 1.2(--09) & 5.7(--06) & 4.4(--07) & 3.8(+02) & EO \\
s-C$_6$H$_6$      & 6.6(--12) & 3.0(--19) & 9.3(--08) & 1.4(+04) & I  \\
H$_2$CO           & 1.8(--11) & 8.7(--07) & 5.8(--08) & 3.3(+03) & EO \\
HCN               & 2.2(--12) & 1.7(--07) & 1.1(--08) & 5.1(+03) & EO \\
s-C$_2$H$_5$CN    & 1.0(--11) & 1.3(--22) & 9.0(--09) & 8.6(+02) & I  \\
S                 & 4.2(--11) & 8.0(--11) & 7.0(--09) & 1.7(+02) & I  \\
HCO               & 1.3(--11) & 6.9(--09) & 4.5(--09) & 3.3(+02) & EO \\
s-H$_3$C$_5$N     & 4.8(--12) & 2.4(--22) & 1.2(--09) & 2.5(+02) & I  \\
CN                & 1.4(--12) & 5.7(--10) & 8.8(--10) & 6.3(+02) & EO \\
HNC               & 9.9(--13) & 6.4(--09) & 6.9(--10) & 6.9(+02) & EO \\
s-C$_4$S          & 8.3(--13) & 2.1(--25) & 5.5(--10) & 6.6(+02) & I  \\
s-CH$_3$NH$_2$    & 9.6(--14) & 1.9(--21) & 4.9(--10) & 5.1(+03) & I  \\
s-HCSi            & 2.0(--12) & 1.1(--21) & 4.8(--10) & 2.5(+02) & I  \\
C$_2$H$_4$        & 3.3(--13) & 6.6(--10) & 3.6(--10) & 1.1(+03) & EO \\
H$_3$CO$^+$       & 1.7(--13) & 3.1(--10) & 1.2(--10) & 7.1(+02) & EO \\
s-HCOOCH$_3$      & 2.0(--13) & 4.9(--13) & 1.2(--10) & 6.0(+02) & EO \\
s-CH$_3$C$_6$H    & 5.1(--14) & 4.1(--12) & 1.1(--10) & 2.1(+03) & IM \\
Cl                & 8.2(--14) & 9.8(--10) & 9.8(--11) & 1.2(+03) & EO \\
C$_2$H$_2$        & 2.3(--13) & 5.7(--11) & 7.1(--11) & 3.2(+02) & EO \\
s-HCOOCH$_3$      & 1.2(--13) & 5.0(--11) & 6.0(--11) & 4.9(+02) & O  \\
P                 & 3.3(--14) & 2.7(--12) & 5.5(--11) & 1.6(+03) & I  \\
NS                & 3.8(--14) & 3.7(--10) & 4.8(--11) & 1.3(+03) & EO \\
HCOOCH$_3$        & 3.1(--18) & 1.2(--12) & 4.4(--11) & 1.4(+07) & EO \\
H$_2$CN$^+$       & 1.0(--13) & 2.1(--11) & 3.9(--11) & 3.8(+02) & EO \\
s-CH$_3$CN        & 1.6(--13) & 2.5(--25) & 3.6(--11) & 2.2(+02) & I  \\
C                 & 1.9(--13) & 8.8(--12) & 3.3(--11) & 1.7(+02) & EO \\
C$_2$H            & 7.7(--14) & 1.7(--11) & 2.8(--11) & 3.6(+02) & EO \\
CH$_2$CN          & 4.7(--15) & 1.7(--11) & 2.0(--11) & 4.3(+03) & EO \\
s-NH$_2$OH        & 3.4(--27) & 1.7(--11) & 1.7(--11) & 4.9(+15) & O  \\
P                 & 2.0(--16) & 1.9(--10) & 1.6(--11) & 8.2(+04) & EO \\
NH$_2$CHO         & 4.3(--16) & 9.8(--12) & 1.4(--11) & 3.3(+04) & EO \\
CH$_3$CN          & 4.1(--15) & 1.7(--11) & 1.4(--11) & 3.5(+03) & EO \\
OCN               & 2.3(--15) & 8.5(--13) & 1.2(--11) & 5.1(+03) & EO \\
C$_2$H$_3$        & 8.9(--14) & 1.6(--11) & 1.0(--11) & 1.2(+02) & EO \\
CH$_2$CO          & 3.1(--17) & 5.8(--11) & 7.3(--12) & 2.3(+05) & EO \\
C$_3$H$_3$N       & 1.7(--15) & 3.5(--12) & 6.9(--12) & 4.0(+03) & EO \\
C$_2$             & 3.8(--15) & 2.5(--12) & 6.9(--12) & 1.8(+03) & EO \\
C$_4$S            & 8.7(--15) & 8.6(--13) & 5.7(--12) & 6.6(+02) & I  \\
OCS               & 1.3(--14) & 2.1(--12) & 4.7(--12) & 3.6(+02) & EO \\
HCP               & 3.7(--14) & 3.2(--14) & 4.7(--12) & 1.3(+02) & I  \\
O$_2$             & 2.1(--15) & 4.7(--11) & 4.1(--12) & 1.9(+03) & EO \\
s-HC$_3$N         & 7.8(--15) & 3.1(--21) & 3.9(--12) & 5.0(+02) & I  \\
CO$_2$            & 1.9(--15) & 4.7(--13) & 3.9(--12) & 2.1(+03) & EO \\
H$_5$C$_2$O$_2^+$ & 1.8(--18) & 2.2(--12) & 3.7(--12) & 2.0(+06) & EO \\
s-HC$_2$NC        & 4.9(--16) & 1.2(--24) & 2.7(--12) & 5.5(+03) & I  \\
H$_2$CN           & 5.3(--15) & 8.4(--14) & 2.3(--12) & 4.4(+02) & EO \\
HC$_3$N           & 1.2(--15) & 1.0(--12) & 2.2(--12) & 1.8(+03) & EO \\
s-CH$_3$OH        & 1.1(--15) & 5.7(--24) & 2.1(--12) & 1.9(+03) & I  \\
HCl               & 1.3(--16) & 1.8(--11) & 2.0(--12) & 1.5(+04) & EO \\
C$_2$O            & 1.8(--17) & 1.2(--13) & 1.7(--12) & 9.5(+04) & EO \\
CH$_3$CHO         & 5.1(--17) & 6.8(--13) & 1.3(--12) & 2.5(+04) & EO \\
C                 & 7.8(--15) & 3.0(--15) & 1.3(--12) & 1.6(+02) & I  \\
CN$^-$            & 6.1(--17) & 6.9(--13) & 1.2(--12) & 1.9(+04) & EO \\
\hline
\end{tabular}
\end{table}

\begin{figure}
\includegraphics[width=\columnwidth]{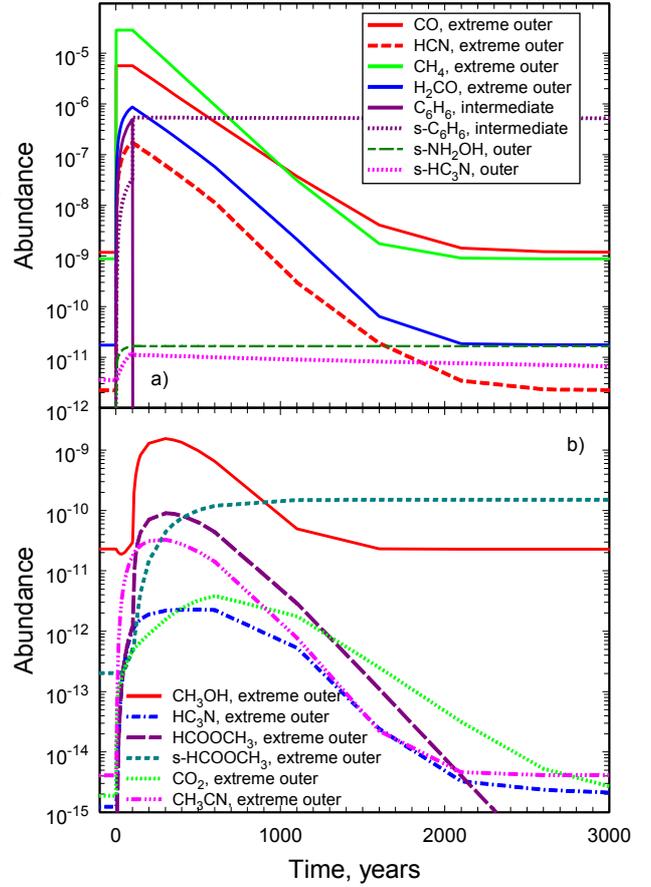}
    \caption{Evolution of some abundances of the species of the third type in various dark disc regions through the outburst. The outburst starts at $t=0$ years and lasts for 100 years. Surface species are indicated with `s-'. Boldfaced are abundances of species that are destroyed by the outburst.}
    \label{wiebe_fig04}
\end{figure}

One mode of keeping the chemical memory of the outburst for a long time is demonstrated by simple molecules like CO and CH$_4$ as shown in Fig.~\ref{wiebe_fig04}a. At the disc periphery, these molecules are abundant ice constituents prior to the outburst. During the outburst their abundances jump up, and after the outburst they re-freeze onto grain surfaces, but because of low density the freeze-out process proceeds slower. The pre-outburst abundance of s-CO and s-CH$_4$ is only restored $\sim3000$ yr after the end of the outburst. Even after several hundred years of the post-outburst evolution CO abundance is still two orders of magnitude higher than the pre-outburst level. Note, however, that the CO evolution of this kind is only seen at the extreme outer disc. This implies, in particular, that CO can be used as a reliable tracer of the past outburst activity only at relatively large distances from the star. A similar behaviour is also demonstrated by other small species like HCN and formaldehyde, also shown in Fig.~\ref{wiebe_fig04}a. Thus, CO, CH4, and HCN can be used as outburst tracers in the outer regions of the discs and external envelopes.

Another mode is characteristic of other, less abundant molecules (Fig.~\ref{wiebe_fig04}b). Its striking feature is that abundances of these species grow {\em after\/} the end of the outburst. For instance, the gas-phase abundance of methanol at the extreme outer disc slightly drops after the onset of the outburst, the molecule being destroyed by some ions (like HCO$^+$), which become abundant during the outburst. At the same time, due to higher dust temperature, desorption of hydrogen atoms intensifies, and they move from dust to gas phase. At the end of the outburst, H gas-phase abundance grows by a factor of several. When luminosity returns to the pre-outburst level, surface methanol synthesis intensifies due to the return of H atoms on grain surfaces, and, accordingly, the gas-phase methanol abundance grows due to reactive desorption. In about 200 years after the end of the outburst it reaches the value of $1.6\times10^{-9}$, which is 50 times higher than the gas-phase methanol abundance at the end of the quiescent stage, and then slowly decreases. Other organic compounds, like CH$_3$CN, HC$_3$N and methyl formate (HCOOCH$_3$), demonstrate similar trends. Their post-outburst abundances, while being somewhat lower than that of methanol, exceed the corresponding pre-outburst abundances by many orders of magnitude. As the pre-outburst abundances are very low, the mere presence of these species in the dark disc gas-phase is an indication of the outburst, which has happened a few hundred years ago.

Let us consider the case of gas-phase methyl formate in more detail. Its main production route at the extreme outer dark disc is dissociative recombination of protonated methyl formate (H$_5$C$_2$O$_2^+$). The abundance of the latter increases after the outburst due to a \hbox{H$_3$CO$^+$ + H$_2$CO} reaction\footnote{The reaction between H$_3$CO$^+$ and H$_2$CO has been added to the ALCHEMIC network from the version of the original OSU network used in \citet{2010A&A...522A..42S}, which is available at \url{http://kida.obs.u-bordeaux1.fr/uploads/models/benchmark_2010.dat}. Its parameters are close to those determined in \citet{2004ApJ...611..605H}. If there were no this reaction, the behaviour of methyl formate would be completely different.}. This reaction is characterized by an unusual $\beta=-3$ value, which implies that the rate coefficient of this reaction increases sharply, when the temperature drops after the end of the outburst. At the same time, the reactants are effectively synthesized during the outburst (they both belong to the third type at this location). Intense gas-phase methyl formate synthesis after the outburst also leads to its accumulation on dust surfaces. The abundance of s-HCOOCH$_3$ stays high even by the end of the computational time ($10^4$ years after the end of the outburst).

The largest ratio of the post-outburst abundance to the pre-outburst abundance ($4.8\times10^{15}$) is obtained for surface hydroxylamine (s-NH$_2$OH) in the outer disc. While the abundances of most species return to their pre-outburst values after a few hundred years of the post-outburst evolution, surface hydroxylamine is an outstanding exception. Its main production route is the reaction between s-NH$_2$ and s-OH, which is suppressed at the pre-outburst stage in favour of the s-NH$_2$~+~s-H reaction, leading to ammonia synthesis. During the outburst the surface abundance of atomic hydrogen drops significantly due to its volatility, and reactions between heavier species start to dominate, including hydroxylamine formation. It rapidly accumulates on dust surfaces and stays there after the outburst ends (Fig.~\ref{wiebe_fig04}a), as it is most effectively destroyed by XCR-induced photons, and the rate of this process is small at this location. To a somewhat less degree and on a shorter time-scale, s-HC$_3$N shows the same behaviour (Fig.~\ref{wiebe_fig04}a). We note the huge ratio of the post-outburst and pre-outburst abundances for hydroxylamine is not physically meaningful as its pre-outburst abundance of hydroxylamine is close to numerical uncertainty. But still, this small abundance indicates that this molecule is unobservable in any practical sense at the quiescent stage, and this is what makes it the most obvious outburst tracer. It should be perfectly invisible in the quiescent disc, so that the very fact of its presence is an indication of the past luminosity outburst.

In the inner and intermediate dark disc locations, interesting cases are represented by benzene (s-C$_6$H$_6$) and methylamine (s-CH$_3$NH$_2$) ices (Fig.~\ref{wiebe_fig04}a). The gas-phase benzene belongs to the second type at these locations as it is effectively synthesized during the outburst, but in fact the benzene synthesis does not stop after the outburst has ended. The major C$_6$H$_6$ production route is the dissociative recombination of C$_6$H$_7^+$, which is moderated during the outburst by the competitive reaction of this ion with ammonia. At the post-outburst stage, ammonia freezes out almost immediately, and benzene starts to be produced in larger amounts. However, all the newly synthesized benzene molecules stick to dust grains, and C$_6$H$_6$ surface abundance at the inner disc 500 years after the outburst is more than four orders of magnitude higher than its pre-outburst abundance.

A post-outburst overabundance of s-CH$_3$NH$_2$ in the inner disc is related to methanimine (CH$_2$NH) evolution. This molecule is effectively produced on warmed dust surfaces and then released from mantles during the outburst. After the end of the outburst this molecule no longer evaporates off the dust grains and rapidly reacts with hydrogen atoms, which are also abundant at the beginning of the post-outburst stage. The resultant chain of hydrogen addition reactions
\[
\mbox{s-CH}_2\mbox{NH}\rightarrow _{\mbox{s-CH}_3\mbox{NH}}^{\mbox{s-CH}_2\mbox{NH}_2}\rightarrow \mbox{s-CH}_3\mbox{NH}_2
\]
leads to formation of surface methylamine, which is a dead end of the adopted surface reaction network. An enhanced post-outburst abundance of propionitrile ice (s-C$_2$H$_5$CN) appears for similar reasons. The intense gas-phase synthesis of HC$_3$N during the outburst is followed by its freeze-out and conversion to s-C$_2$H$_5$CN by consecutive surface hydrogen addition reactions.

While surface species with enhanced post-outburst abundances are not directly observable (if there is no suitable infrared background source), they may become visible, if the next (recurrent) outburst happens within less than $\sim500$ years from the previous one. Such clustered outbursts were predicted in the disc fragmentation model of \citet{2015ApJ...805..115V}.

The list in Table~\ref{ddgr3} also includes some species with less abundant atoms, including atomic phosphorus. Their appearance in the post-outburst disc is related to chemical reactions with some leftovers of the outburst chemistry, in particular, with HCO$^+$ and acetylene (C$_2$H$_2$). Specifically, the enhanced P abundance is related to PO destruction in the reaction with HCO$^+$.

\subsection{Atmosphere}

An additional factor that we introduce in the atmospheric locations is the rise in UV illumination of the disc material. As we mentioned earlier, taking into account a response of the disc atmosphere to the luminosity outburst is less trivial than in the case of the disc dark regions because of the outburst-driven structure changes in the upper disc layers. Nevertheless, we think that it is worth considering the chemical effect of the enhanced UV irradiation. We follow the same procedure as in Section 3.1. The chemical evolution is simulated for the four disc radii (inner, intermediate, outer, and extreme outer), and then all the species are divided into four behaviour types.

\subsubsection{First type}

\begin{figure}
\includegraphics[width=\columnwidth]{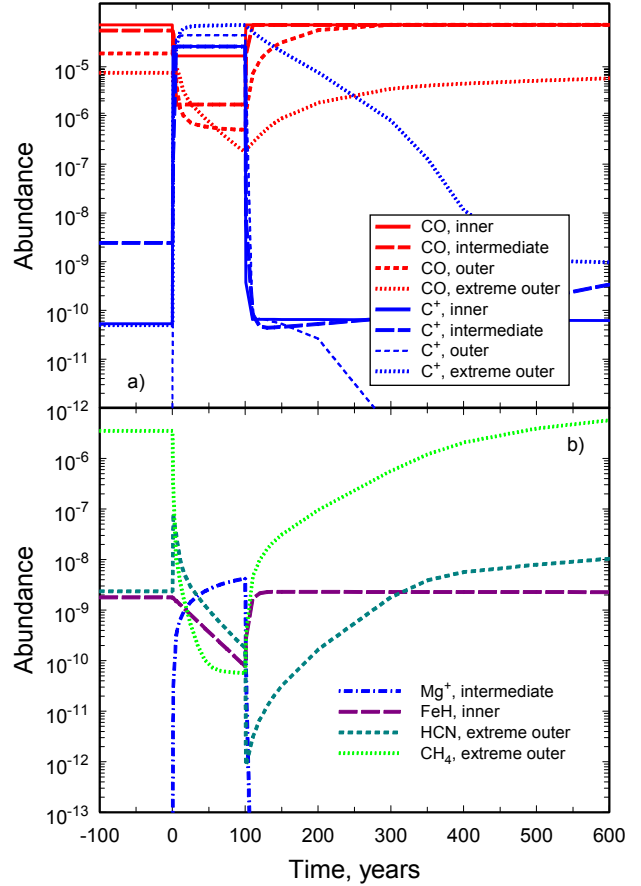}
    \caption{Evolution of some abundances of the species of the first and second types in the disc atmosphere through the outburst. The outburst starts at $t=0$ years and lasts for 100 years.}
    \label{wiebe_fig05}
\end{figure}

Quite expectedly, behaviour of the first type is mostly demonstrated by mono-atomic ions and neutral atoms as shown in Fig.~\ref{wiebe_fig05}a and Table~\ref{agr1}. The most abundant ion during the outburst is C$^+$. Its abundance jumps up by almost eight orders of magnitude, when the luminosity increases. Abundances of Si$^+$, Mg$^+$, and Na$^+$ grow even stronger, but still stay low because of the low total content adopted for these elements. Molecules mostly show upturned $\Pi$-like profiles, being photodissociated during the outburst (CO is shown as an example in Fig.~\ref{wiebe_fig05}a); their $x_{\rm o}/x_{\rm q}$ ratios are boldfaced in Table~\ref{agr1}.

\begin{table}
\caption{Abundances of species of the first type in the atmosphere, sorted by the decreasing $x_{\rm o}$ value. In the last column a location code is indicated: I --- inner disc, IM --- intermediate disc, O --- outer disc.}
\label{agr1}
\begin{tabular}{llllll}
\hline
Species    & $x_{\rm q}$   & $x_{\rm o}$ &  $x_{\rm p}$ & $R_{\rm oq}$ & Loca-\\
 &   &  &   & & tion\\
\hline
H$_2$  &   2.8(--01)  &  4.4(--04)  &  5.1(--02) &   {\bf 1.6(--03)}  &       I  \\
C$^+$  &   5.3(--13)  &  4.4(--05)  &  1.9(--13) &   8.4(+07)         &       O  \\ 
C      &   2.8(--11)  &  3.0(--05)  &  2.7(--11) &   1.1(+06)         &       I  \\
C$^+$  &   5.3(--11)  &  2.6(--05)  &  6.2(--11) &   4.9(+05)         &       I  \\
C$^+$  &   2.4(--09)  &  2.6(--05)  &  3.4(--10) &   1.1(+04)         &       IM \\
N      &   1.9(--08)  &  2.1(--05)  &  5.0(--08) &   1.1(+03)         &       I  \\
N      &   1.9(--07)  &  2.1(--05)  &  1.9(--07) &   1.1(+02)         &       IM \\
CO     &   5.5(--05)  &  1.7(--06)  &  7.2(--05) &   {\bf 3.0(--02)}  &       IM \\
H$^+$  &   3.3(--10)  &  1.5(--06)  &  6.2(--10) &   4.5(+03)         &       I  \\
H$^+$  &   4.8(--09)  &  1.2(--07)  &  2.5(--09) &   2.5(+01)         &       IM \\
S$^+$  &   7.8(--12)  &  7.9(--08)  &  9.9(--12) &   1.0(+04)         &       I  \\
S$^+$  &   2.0(--09)  &  7.9(--08)  &  4.8(--10) &   3.9(+01)         &       IM \\
O$_2$  &   3.7(--05)  &  8.2(--09)  &  3.7(--05) &   {\bf 2.2(--04)}  &       I  \\
Si$^+$ &   9.7(--13)  &  7.9(--09)  &  2.4(--12) &   8.2(+03)         &       I  \\
Si$^+$ &   2.1(--18)  &  7.9(--09)  &  1.6(--18) &   3.7(+09)         &       O  \\
\hline
\end{tabular}
\end{table}

In this case we also do not see any species of the first type in the extreme outer disc. Note that ionized silicon only belongs to the first type in the inner and outer disc. In fact, its behaviour in the intermediate and extreme outer disc is nearly the same, but the restoration of the quiescent abundance proceeds somewhat slower at these locations, so that Si$^+$ behaviour marginally falls into the third type.

\subsubsection{Second type}

Among the species, whose abundances are high and evolving at the outburst stage, only four, Mg, Mg$^+$ (Fig.~\ref{wiebe_fig05}b), Fe, and Fe$^+$, have their abundances increased due to the flare. Other typical patterns are also shown in Fig.~\ref{wiebe_fig05}b. Species that are mostly contained in the gas phase at the quiescent stage, like methane, are dissociated by photons during the outburst. At large distances from the star, this process is quite slow, and the abundances of such species continue to show a noticeable decrease for the entire outburst duration. Abundances of species that are mostly contained in the solid phase at the quiescent stage, like HCN, first grow due to thermal desorption and photodesorption and then decrease due to photodissociation. An interesting behaviour is demonstrated by FeH in the inner disc. At the quiescent stage, this molecule is synthesized on dust surfaces (with subsequent desorption) and destroyed by XCR-induced photons. During the outburst the surface synthesis is quenched due to high dust temperature, and the balance is thus broken in favour of destruction, so the abundance of FeH decreases linearly with time. The effect of the outburst onto this molecule would be more significant (in the considered setup), should the used network also include the dissociation of FeH by `ordinary' photons. But for some reasons this reaction is missing in publicly available networks.

\subsubsection{Third type}

\begin{figure}
\includegraphics[width=\columnwidth,clip=]{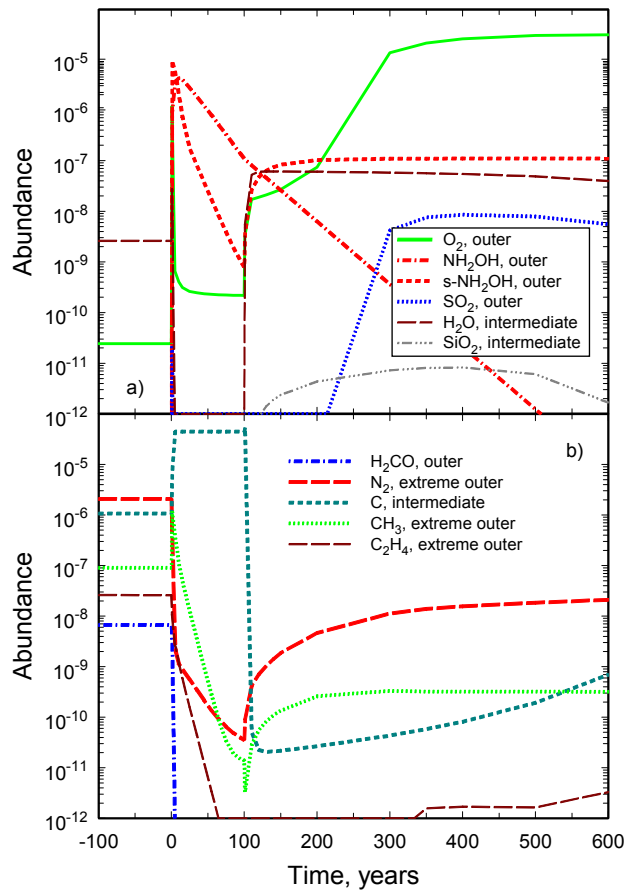}
    \caption{Evolution of some abundances of the species of the third and fourth types in the disc atmosphere through the outburst. The outburst starts at $t=0$ years and lasts for 100 years.}
    \label{fig6}
\end{figure}

Species of the third type are mostly present in the outer and extreme outer atmosphere locations (see Table~\ref{agr3} and Fig.~\ref{fig6}a). Prior to the outburst, nearly all O atoms are bound in water ice and gas-phase CO. During the outburst both molecules are photo-dissociated, making oxygen atoms available for other molecules, most notably for O$_2$. Reactions of O$_2$ with sulfur and silicon cause enhanced abundances of SO, SO$_2$ and SiO$_2$ (Table~\ref{agr3}), which are retained in the gas-phase for a relatively long time interval. Then, O atoms are gradually transferred back to CO and water ice. The growth of sulfur-bearing abundances is interesting, as these components are proposed as tracers of dense fragments in protoplanetary discs \citep[see e.g.][]{2017MNRAS.472..189I}.

The surface hydroxylamine (s-NH$_2$OH) abundance jumps at the start of the outburst for the reasons indicated above, but then within a year it is photoevaporated, and its later evolution in the gas phase is governed by slow XCR-induced photodissociation and freeze-out (after the end of the outburst; see Fig.~\ref{fig6}a). This molecule also shows a dramatic increase of the gas-phase abundance during and after the outburst, over 10 orders of magnitude in the outer disc atmosphere.

\begin{table}
\caption{Abundances of species of the third type in the atmosphere, sorted by the decreasing $R_{\rm pq}$ ratio. In the last column a location code is indicated: IM --- intermediate disc, O --- outer disc, EO --- extreme outer disc.}
\label{agr3}
\begin{tabular}{llllll}
\hline
Species    & $x_{\rm q}$   & $x_{\rm o}$ &  $x_{\rm p}$ & $R_{\rm pq}$ & Loca-\\
 &   &  &   & & tion\\
\hline
NH$_2$OH   &3.5(--24) &1.1(--07) &7.2(--14) &2.0(+10) &O  \\
s-NH$_2$OH &6.4(--18) &7.7(--10) &1.1(--07) &1.7(+10) &O  \\
CHNH       &1.3(--20) &2.3(--11) &2.0(--11) &1.5(+09) &EO \\
O$_3$      &1.5(--20) &8.5(--25) &5.8(--12) &3.9(+08) &O  \\
O$_3$      &3.4(--21) &1.4(--14) &4.1(--13) &1.2(+08) &EO \\
O$_2^+$    &3.5(--18) &1.9(--14) &7.5(--11) &2.2(+07) &O  \\
SO$_2$     &8.1(--16) &5.1(--22) &5.5(--09) &6.8(+06) &O  \\
OCS$^+$    &4.8(--18) &4.4(--21) &1.9(--11) &3.9(+06) &EO \\
S$_2^+$    &1.8(--18) &1.2(--25) &3.0(--12) &1.7(+06) &O  \\
O$_2$      &2.4(--11) &2.2(--10) &3.1(--05) &1.3(+06) &O  \\
Si         &5.5(--15) &2.6(--11) &6.4(--09) &1.2(+06) &EO \\
\multicolumn{6}{l}{...} \\
SiO$_2$    &2.3(--17) &7.3(--22) &1.7(--12) &7.4(+04) &IM \\
\multicolumn{6}{l}{...} \\
H$_2$O     &2.6(--09) &3.9(--13) &4.0(--08) &1.5(+01) &IM \\
\hline
\end{tabular}
\end{table}

Note that nearly all molecules capable of retaining the outburst memory are located in the outer and extreme outer disc regions. Only few molecules keep the memory of the outburst closer to the star; two examples (SiO$_2$ and H$_2$O) are shown in the bottom of Table~\ref{agr3} and  in Fig.~\ref{fig6}a. In both cases, the post-outburst evolution is related to the CO photodissociation and re-distribution of oxygen atoms between gas-phase CO, water, SiO$_2$, and other molecules. Their pre-outburst abundances are only recovered after several centuries.

\subsubsection{Fourth type}

In Fig.~\ref{fig6}b we show evolution of abundances for some gas-phase species that are more or less abundant at the quiescent stage, but are destroyed by the outburst, and do not reappear in the disc atmosphere for at least 500 years after the outburst. These are mostly species that are photodissociated during the outburst and then are restored slowly after its end. The important exception is atomic carbon. While its abundance grows during the outburst, after the outburst atomic carbon actually becomes less abundant than at the quiescent stage due to reactions with abundant O$_2$ molecules.

\section{Wandering away from the fiducial model}

In this section, we make digressions from the fiducial model to probe the sensitivity of our conclusions to variations in the model parameters. Again, we present results in terms of the ratios of the outburst abundances (by the end of the outburst) to that at the immediate pre-outburst stage ($R_{\rm oq}=x_{\rm o}/x_{\rm q}$) and the ratios of the post-outburst abundances at $t=500$ yr after the end of the outburst to that at the immediate pre-outburst stage ($R_{\rm pq}=x_{\rm p}/x_{\rm q}$). The first ratio shows, which species are most responsive to the outburst itself (types 1 and 2), while the second ratio reveals species, which are capable of retaining memory of the outburst for at least a few centuries (type 3). As the chemical response to the outburst in the atmosphere is not that diverse, in this section we only show results for dark disc locations.

\subsection{Dust size distribution}

The most obvious element of the model subject to variations is the dust size distribution. In this work we assume the MRN-like power law distribution with a fixed lower size limit $a_{\min}=5\cdot10^{-7}$~cm, a power-law index $p=-3.5$, and variable upper size limit $a_{\max}$. In the fiducial model, $a_{\max}$ is taken to be 1\,$\mu$m. In this subsection, we present results for larger values of $a_{\max}$. In the adopted formalism, similar to the one used in \cite{2017A&A...604A..15R}, an order of magnitude increase of $a_{\max}$ leads to an order of magnitude decrease in the grain surface area available to solid-phase reactions. Note that we only use an increased value of $a_{\max}$ at stages corresponding to the disc evolution. The molecular cloud stage is always simulated with $a_{\max}$ of 1\,$\mu$m. When passing from the molecular cloud stage to the pre-outburst disc stage, we assume that surface species are transferred from `cloud' grains to `disc' grains irrespective of the changes in the total dust surface area. Physically, this means that in models with increased $a_{\max}$ icy mantles are assumed to be preserved entirely during grain coagulation.

Quite naturally, the effect of dust size variations is most noticeable in abundances of species that either reside on grains or depend significantly on surface reactions. For example, in the inner dark disc the sensitivity of gas-phase molecules experiencing evaporation from the grain mantles as a major outburst effect weakens as the dust surface area becomes smaller. In the fiducial model, many species residing on dust surfaces during the quiescent period are characterized by quite significant values of $R_{\rm oq}$, $10^{10}$ or greater. In models with increased $a_{\max}$ their $R_{\rm oq}$ decrease by orders of magnitude, but still remain quite high. So, in the inner disc even $a_{\max}$ as high as 10 cm does not alter significantly the distribution of molecules over sensitivity types.

\begin{figure}
\includegraphics[width=\columnwidth]{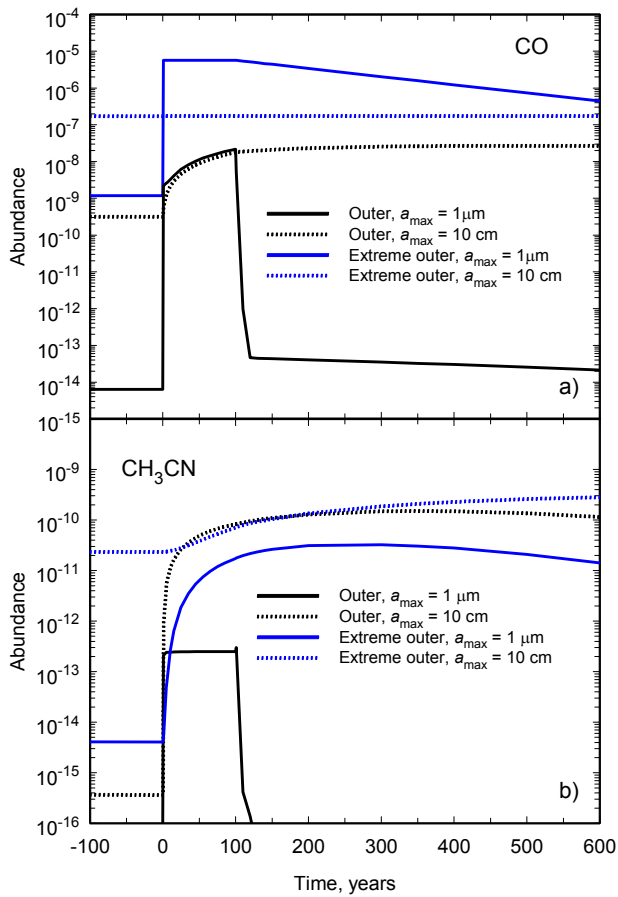}
    \caption{Evolution of CO and CH$_3$CN abundances in models with fiducial and increased $a_{\max}$ values.}
    \label{fig07}
\end{figure}

The situation is different in the outer and extreme outer disc. In Fig.~\ref{fig07} we compare the evolution of CO and CH$_3$CN in the outer and extreme outer disc locations for the fiducial model and the model with $a_{\max}=10$~cm. Both molecules (along with many other organics, for example, carbon chains) share the same behaviour. At the outer disc, they belong to the type~1 in the fiducial model (solid black lines), but move into the type~3 in the model with grown dust (dotted black lines) as the reduced dust surface slows down the re-freezing process significantly. However, at the extreme outer dark disc, they move from the type~3 to type~0 (solid and dotted blue lines), in the sense that lower density and lower dust surface cause these molecules to stay in the gas phase before the outburst, so that it does not affect their abundances significantly. In other words, as dust grows, many organic molecules become sensitive outburst tracers at about 100~au from the star, but cease to be its tracers farther out in the disc. Whether or not this change can be observationally detectable depends on the mass distribution in the disc.

\begin{figure}
\includegraphics[width=\columnwidth]{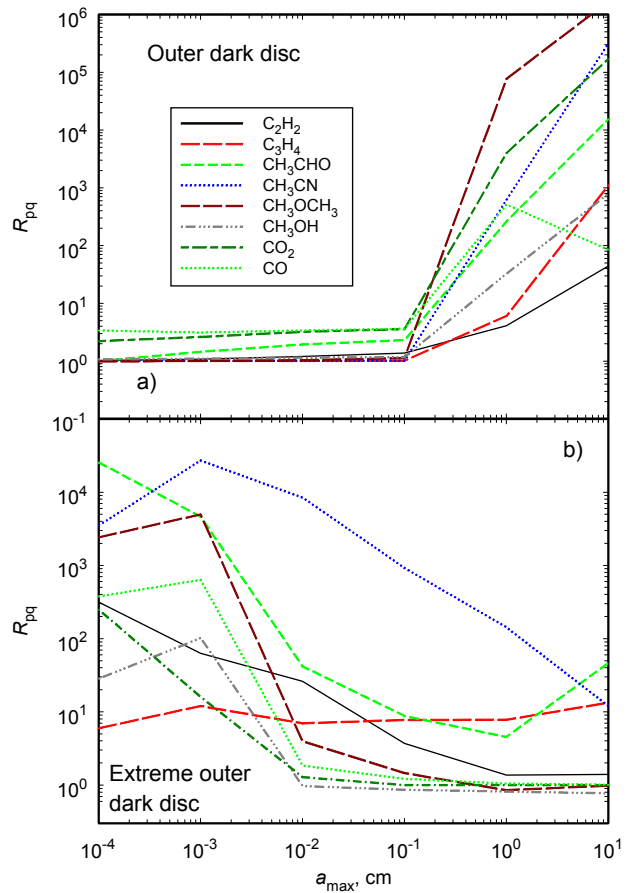}
    \caption{Abundance ratios $R_{\rm pq}$ for some carbon-bearing molecules as functions of the maximum grain size, $a_{\max}$. Same legend applies to both panels.}
    \label{fig08}
\end{figure}

This is further illustrated by Fig.~\ref{fig08}, where we plot abundance ratios $R_{\rm pq}$ for some carbon-bearing molecules as functions of the maximum grain size for the outer (top panel) and extreme outer (bottom panel) disc locations. Opposite trends are seen in these two locations. While at the outer disc all the plotted species become sensitive late outburst tracers as the dust maximum size becomes greater than 0.1~cm, at the extreme outer disc their sensitivity weakens in models with increased $a_{\max}$ values. Only a few molecules (like propyne, C$_3$H$_4$, acetaldehyde, CH$_3$CHO, and acetonitrile, CH$_3$CN, shown in Fig.~\ref{fig08}) become sensitive to past outbursts in models with high $a_{\max}$ at the outer disc, while retaining this sensitivity at the extreme outer disc.

\begin{table}
\caption{Species that belong to the type~3 either in the model with $a_{\max}=10$~cm or in the fiducial model.}
\label{nowyouseemenowyoudont}
\begin{tabular}{lllll}
\hline
\multicolumn{5}{c}{Type~3 in the model with $a_{\max}=10$~cm only}\\
\multicolumn{5}{l}{\em Inner disc:}\\
H$_2$SiO   & s-H$_2$SiO   & s-H$_3$C$_5$N & NaOH         & NH \\
NH$_2$     & s-NH$_2$CHO  & S$^+$         & Si           & SiH$_4$ \\
s-SiH$_4$  & SiN          &               &              & \\
\multicolumn{5}{l}{\em Outer disc:}\\
C$_2$         & C$_2$H$_2$   & s-C$_2$H$_2$  & C$_2$H$_3$   & C$_2$H$_4$        \\
C$_2$H$_6$    & C$_2$S       & C$_3$H$_2$    & s-C$_3$H$_2$ & C$_3$H$_3$        \\
C$_3$H$_3$N   & s-C$_3$H$_3$N& C$_3$H$_4$    & C$_3$S       & C$_4$H            \\
C$_4$H$_2$    & C$_4$N       & C$_5$H$_2$    & C$_6$H$_6$   & CH$_2$CN          \\
CH$_2$CO      & CH$_2$OH     & CH$_3$CHO     & CH$_3$CN     & CH$_3$CO$^+$      \\
CH$_3$OCH$_3$ & s-CH$_2$CN   & CH$_3$OH      & CH$_4$       & Cl           \\
CO            & CO$_2$       & s-CO          & CS           & s-CS         \\
H$_2$C$_3$O   & H$_2$CO      & H$_2$CS       & H$_2$S$_2$   & H$_3$CO$^+$  \\
H$_3$CS$^+$   & H$_5$C$_2$O$_2^+$ & HC$_3$N  & s-HC$_3$N    & HC$_3$O       \\
HC$_5$N       & HCl          & HCN           & HCOOCH$_3$   & HCOOH          \\
HCS           & HNC          & HNCO          & HNO          & HS$_2$           \\
N$_2$O        & NH$_2$CHO    & NH$_2$OH      & OCN          & OCS                  \\
P             & PH           & PN            & PO           & S$_2$ \\
SiC           & SiC$_2$      & SiC$_2$H      & SiCH$_2$     & SiH$_4$ \\
SiN           & SiO$_2$      & SiO           & SO &\\
\multicolumn{5}{l}{\em Extreme outer disc:}\\
C$_3$H$_4$    & C$_4$H$_3$   & C$_4$H$_5^+$  & C$_5$H$_4^+$ & C$_5$H$_5^+$ \\
C$_9$H$_4^+$  & s-CH$_3$CHO  & SiN           &         &\\
\hline
\multicolumn{5}{c}{Type~3 in the fiducial model only}\\
\multicolumn{5}{l}{\em Inner disc:}\\
C$_3$H$_2$    & C$_3$O       & s-C$_9$H$_2$  & C$_9$H$_4$   & s-C$_9$H$_4$ \\
CH$_2$NH      & s-CH$_3$C$_6$H & CH$_3$NH$_2$& s-CH$_3$NH$_2$ & H$_5$C$_3$N \\
HC$_2$NC      & s-HC$_2$NC   & HC$_3$N       & s-HC$_3$N    & HCN \\
HCSi          & OCS          & SiC$_2$H      & &\\
\multicolumn{5}{l}{\em Intermediate disc:}\\
s-C$_6$H$_2$  & C$_6$H$_4$   & s-C$_6$H$_4$  & C$_6$H$_5$CN & s-C$_9$H$_2$ \\
C$_9$H$_4$    & s-C$_9$H$_4$ & s-CH$_3$C$_6$H &&\\
\multicolumn{5}{l}{\em Extreme outer disc:}\\
C           & C$_2$       & C$_2$H      & C$_2$H$_2$    & C$_2$O \\
C$_3$H$_2$  & C$_3$H$_3$N & s-C$_3$H$_3$N & CH          & CH$_2$     \\
CH$_2$CN    & CH2CO CH$_3$& CH$_3$OH      & CH$_5^+$    & CN        \\
CN$^-$      & CO          & CO$_2$        & CS          & H$_2$CN     \\
H$_2$CN$^+$ & H$_2$CO     & H$_2$CO$^+$   & H$_2$CS     & H$_3$CO$^+$\\
H$_5$C$_2$O$_2^+$ &HC$_3$N& HCN           & HCO         & HCO$^+$      \\
HCOOCH$_3$  & s-HCOOCH$_3$& HNC           & HNO         & N    \\
N$_2$       & NH$_2$CHO   & O             & O$_2$       & s-O$_2$\\
OCN         & OCS&&&\\
\hline
\end{tabular}
\end{table}

For the purpose of reference, in the top part of Table~\ref{nowyouseemenowyoudont} we list species that do not keep the outburst memory in the fiducial model but start to demonstrate the behaviour of the third type as $a_{\max}$ increases. The bottom part of Table~\ref{nowyouseemenowyoudont} contains species that are good tracers of the past outburst activity in the fiducial model but lose this sensitivity in the model with $a_{\max}=10$~cm. While the exact selection of species may depend on the adopted network, some general trends are obvious. First, changes are minimal closer to the star, and they mostly embrace `exotic' complex molecules. Second, at the outer disc dust growth extends the sensitive species list; however, some of them drop out of the corresponding list at the extreme outer disc. These are volatile species, whose outburst gas-phase abundances in the fiducial model grow mostly due to evaporation of icy mantles. A decrease of the total dust surface area leads to a decrease in the solid-phase abundances of these species (like CO, formaldehyde, or HC$_3$N). In the model with grown dust, their $R_{\rm oq}$ and $R_{\rm pq}$ ratios drop to one, implying nearly constant abundances during and after the outburst. The reason is that at the extreme outer disc in the model with grown dust these species mostly reside in the gas phase even at the pre-outburst stage. Only species with large ice fractions (methane, CH$_3$CHO) retain their possibility to keep memory of the outburst in models with increased $a_{\max}$. For methane, a decrease in $R_{\rm oq}$ and $R_{\rm pq}$ related to the dust growth is quite significant, but even when we assume the maximum grain size of 10 cm, the post-outburst gas-phase methane abundance still exceeds the pre-outburst level by a factor of 300.

Yet another example is atomic phosphorus. Its $R_{\rm pq}$ ratio at the extreme outer disc is 82\,000 in the fiducial model (see Table~\ref{ddgr3}). As dust grows, the ratio gets smaller, but even if $a_{\max}=10$~cm $R_{\rm pq}$ for P is still about 5000. Even after 5000 years the gas-phase abundance of phosphorus exceeds its pre-outburst value by a factor of 4700. This is, to a certain degree, a consequence of the adopted treatment of the mantle evolution. During the molecular cloud stage, nearly all P atoms are frozen on dust surfaces. Setting $a_{\max}$ value to 10~cm, we still assume that these P atoms remain locked in the icy mantles. The outburst-related heating releases these atoms and makes them available for some gas-phase processing. Having been released into the gas phase, phosphorus atoms stay there for a long time due to a low dust surface area. This is accompanied by a large increase of PH abundance.

\subsection{Duration of the pre-outburst stage}

\begin{figure}
\includegraphics[width=\columnwidth,clip=]{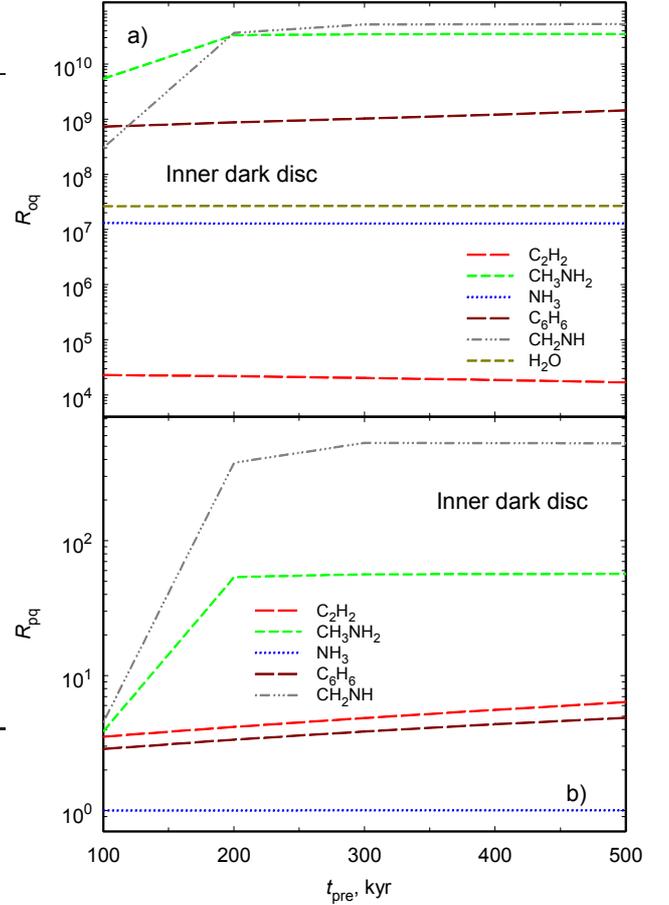}
    \caption{Ratios of abundances $R_{\rm oq}$ and $R_{\rm pq}$ for the inner dark disc model as a function of the pre-outburst stage duration.}
    \label{tpre}
\end{figure}

In the fiducial model, we assume that the disc evolves for 500 kyr in quiescence before the first outburst occurs. This time instance roughly corresponds to the beginning of the T Tauri phase of disc evolution \citep{2009ApJS..181..321E,2010ApJ...713.1059V}. However, the first outburst may actually occur much sooner, already in the embedded phase. In Fig.~\ref{tpre} we present $R_{\rm oq}$ and $R_{\rm pq}$ ratios for the inner dark disc model as functions of the pre-outburst stage duration $t_{\rm pre}$. At this location, both outburst and post-outburst evolution of most species does not depend on the duration of the pre-outburst stage. In the high density and temperature environment, the chemical equilibrium is reached very fast, so that $R_{\rm oq}$ and $R_{\rm pq}$ are nearly the same for almost all species and any $t_{\rm pre}$ that is longer than 100 kyr. An exception is represented by molecules that are accumulated rather slowly, being synthesized in surface reactions followed by reactive desorption, such as methylamine (CH$_3$NH$_2$) and methanimine (CH$_2$NH). Their response to the outburst both in terms of the outburst and post-outburst abundance does not change appreciably for $t_{\rm pre}\gtrsim200$~kyr. In other disc locations farther from the star, the pre-outburst abundances of nearly all molecules do not depend on the duration of the quiescent stage (in the considered limits), because physical parameters at these locations do not differ much from the parameters of the molecular cloud stage.

\subsection{Outburst duration and intensity}

\begin{figure}
\includegraphics[width=\columnwidth,clip=]{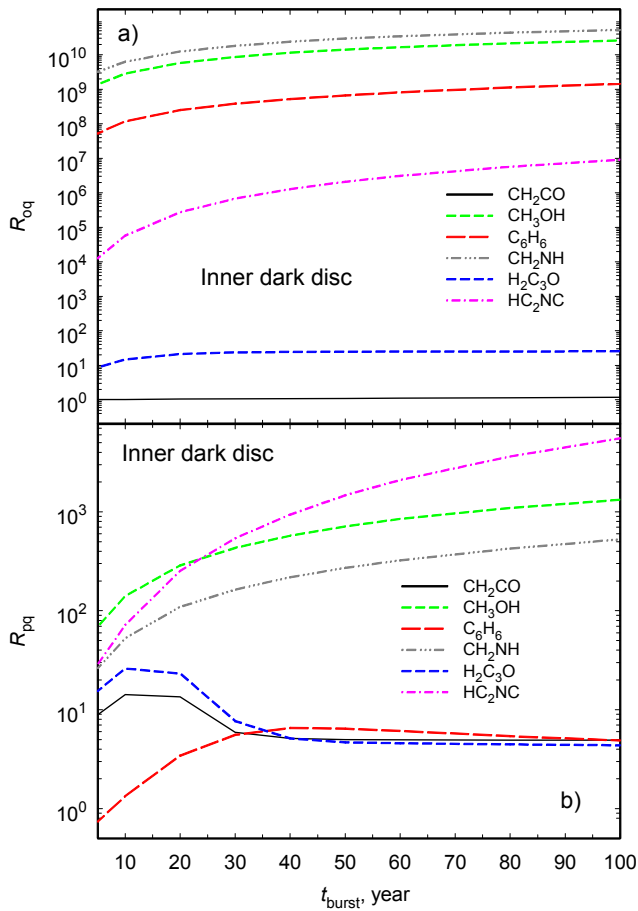}
    \caption{Ratios of abundances $R_{\rm oq}$ and $R_{\rm pq}$ for the inner dark disc model as a function of the outburst duration.}
    \label{tburst1}
\end{figure}

The observationally inferred duration of the outburst phase does not exceed a century and in most cases is limited to several decades \citep{audard}. However, numerical models predict that some outbursts may last longer \citep[e.g.][]{2015ApJ...805..115V}, and several FUors (e.g., FU Orionis itself) show little signs of fading for almost a century. Nevertheless, it is worth considering the effect of a shorter outburst duration on the disc chemical evolution. Fig.~\ref{tburst1} shows the dependence of $R_{\rm oq}$ and $R_{\rm pq}$ parameters on the duration of the outburst, which we vary from 5 year to 100 years, for the inner dark disc. The strength of the outburst (that is, the peak temperature) is kept unaltered.

The outburst abundances of most simple species (like water, CO, CO$_2$) in the inner dark disc are determined by their evaporation from icy mantles, which proceeds quite rapidly. Thus, their abundances during the outburst are nearly independent on its duration. More complex molecules are synthesized during the outburst either in the gas phase or on the dust surfaces and their abundances keep growing during the outburst (Fig.~\ref{tburst1}a). However, the new equilibrium state is established in a few decades, and after that the growth is quite slow. There are only a few exceptions, such as isocyanoacetylene (HC$_2$NC), and we may safely assume that our conclusions on the outburst abundances of most species in the inner dark disc do not strongly depend on the outburst duration if it is greater than 20--30 years.

The situation with the post-outburst abundances is more complicated (Fig.~\ref{tburst1}b) for some O-bearing organic species, like ethenone (CH$_2$CO) and propynal (H$_2$C$_3$O). Let us consider propynal as an example. After the end of the outburst, its abundance first grows by an order of magnitude, reaching a maximum in about 50 years. This growth is caused by the surface processes initiated by the high post-outburst abundances of atomic hydrogen and oxygen. Then the propynal abundance decreases, and the rate of decrease is larger in the model with $t_{\rm burst}=100$ years. The reason is probably hidden in sulfur chemistry. During the outburst, abundances of simple S-bearing species grow (like NS), as at high temperature their formation is more effective than destruction. When the medium returns to its quiescent state, destruction of NS is no longer compensated by its formation, and released sulfur atoms are rapidly transferred to carbon-bearing compounds (CS, C$_2$S, C$_3$S, H$_2$CS). In turn, the XCR-induced photons slowly destroy these compounds, increasing gas-phase C atoms abundance by two orders of magnitude 100 years after the outburst. These atoms turn out to be an important factor of propynal destruction in the reaction C~+~H$_2$C$_3$O~$\rightarrow$~C$_3$H$_2$~+~CO, and after 500 years its abundance diminishes again. This chain of processes is not effective if $t_{\rm burst}<50$ years as in this case the outburst is not long enough to ensure the transfer of some S atoms from SO to NS and then to CS and other species mentioned above.

\begin{figure}
\includegraphics[width=\columnwidth,clip=]{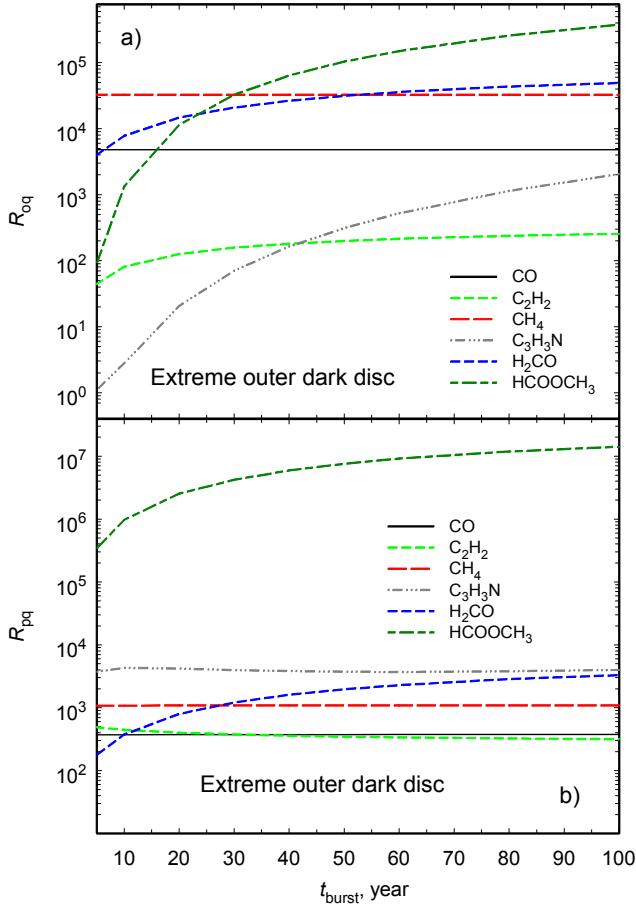}
    \caption{Ratios of abundances $R_{\rm oq}$ and $R_{\rm pq}$ for the extreme outer dark disc model as a function of the outburst duration.}
    \label{tburst2}
\end{figure}

In Fig.~\ref{tburst2} we show $R_{\rm oq}$ and $R_{\rm pq}$ for the extreme outer dark disc in models with varying outburst length. Here we see no unexpected results. Outburst and post-outburst abundances of simple molecules do not depend on the outburst duration, while abundances of more complex species grow with $t_{\rm burst}$ as long as it is smaller than $\sim50$ years. Further increase of $t_{\rm burst}$ does not lead to appreciable changes in $R_{\rm oq}$ and $R_{\rm pq}$.

\begin{figure}
\includegraphics[width=\columnwidth,clip=]{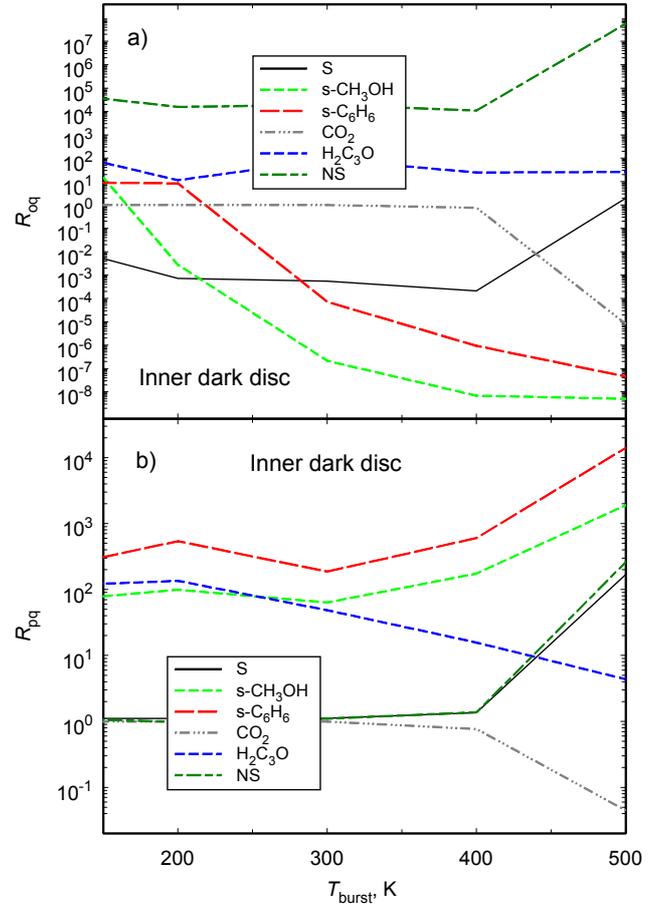}
    \caption{Ratios of abundances $R_{\rm oq}$ and $R_{\rm pq}$ for the inner dark disc model as a function of the outburst intensity.}
    \label{tt1}
\end{figure}

In our model the strength of the outburst is mainly set through the temperature, and in the fiducial model the temperature increase roughly corresponds to the accretion rate of $10^{-5}\,M_{\odot}$ yr$^{-1}$. We also considered a range of outburst temperatures, $T_{\rm burst}$, with the low border corresponding to $\dot{M}\sim10^{-7}\,M_{\odot}$ yr$^{-1}$. Results are shown in Fig.~\ref{tt1}. We do see variations in the behaviour of some abundances, especially, near the upper boundary of the considered temperature range. Outburst gas-phase abundances of S and NS grow significantly, when $T_{\rm burst}$ approaches 500 K, which is important for the post-outburst chemistry (see above). Gas-phase collisional destruction of CO$_2$ is only effective at the upper boundary of the considered $T_{\rm burst}$ range. Also, high post-outburst abundances of methanol and benzene ices are seen in models with high outburst intensities. Results for the outer dark disc are qualitatively similar.

Overall, we conclude that our analysis is valid for any outburst duration exceeding $\sim50$ years and for peak $\dot{M}$ values of the order of $10^{-5}\,M_{\odot}$ yr$^{-1}$. However, shorter and/or less powerful events (EXor-like) in general require a separate analysis.

\subsection{Parameters of the molecular cloud stage}

As we have already noted, our results are to a certain degree shaped by the presence of the molecular cloud stage. An essence of this stage is an accumulation of some less volatile ices that would not have been formed in warm disc regions out of the atomic initial composition. However, due to the adopted chemical model, at least in one case this accumulation is too efficient. This is the case of carbon atoms that are mostly locked in methanol ice by the end of the molecular stage with much less atoms available for other C-bearing ices. This is not what is indicated by available observations that seemingly show more or less equal distribution of C atoms between CO, CO$_2$, and methanol.

There are two ways to change the initial distribution of carbon-bearing ices. One way is to alter parameters of the molecular stage, another way is to introduce some modifications into the used network. In the first case, effect is bound only to the molecular stage, but parameters of this stage can be varied significantly without much violating observational constraints. In the second case, the effect is long-lasting, but the space for manoeuvre seems to be more limited. 

Regarding the parameters of the molecular cloud stage, we have chosen to test two possibilities. The first one is to set a shorter molecular stage duration, $t_{\rm mol}=3\cdot10^5$ yr, and the second one is to raise the temperature during the molecular cloud stage, $T_{\rm mol}$, up to 20\,K. The resultant ice abundances are compared with the ones in the fiducial model in Table~\ref{premol}. In the model with a shorter molecular cloud stage, s-CO does not have enough time to be converted into surface methanol, and the temperature is too low for the effective surface CO$_2$ synthesis. In the model with a higher dust temperature s-CO hydrogenation has to compete with s-CO$_2$ production, so that carbon atoms in ice are distributed between s-CO$_2$ and s-CH$_3$OH molecules. We believe that these two options bracket existing possibilities to a certain degree and can be considered as limiting cases.

\begin{table} 
\caption{Relative ice abundances after the molecular cloud stage.}
\label{premol}
\begin{tabular}{llll}
\hline
Model                       &  s-CO/s-H$_2$O & s-CO$_2$/s-H$_2$O & s-CH$_3$OH/s-H$_2$O\\
Fiducial                    &  1(--04)       & 1(--03)   & 3.2(--01)\\
$t_{\rm mol}=3\cdot10^5$ yr &  2(--01)       & 1(--03)   & 6(--02)\\
$T_{\rm mol}=20$ K          &  9(--05)       & 1.5(--01) & 2.5(--01) \\
Revised                     &  2.5(--01)     & 2(--03)   & 3(--02)\\
network &  &  & \\
\hline
\end{tabular}
\end{table} 

\begin{figure}
\includegraphics[width=\columnwidth]{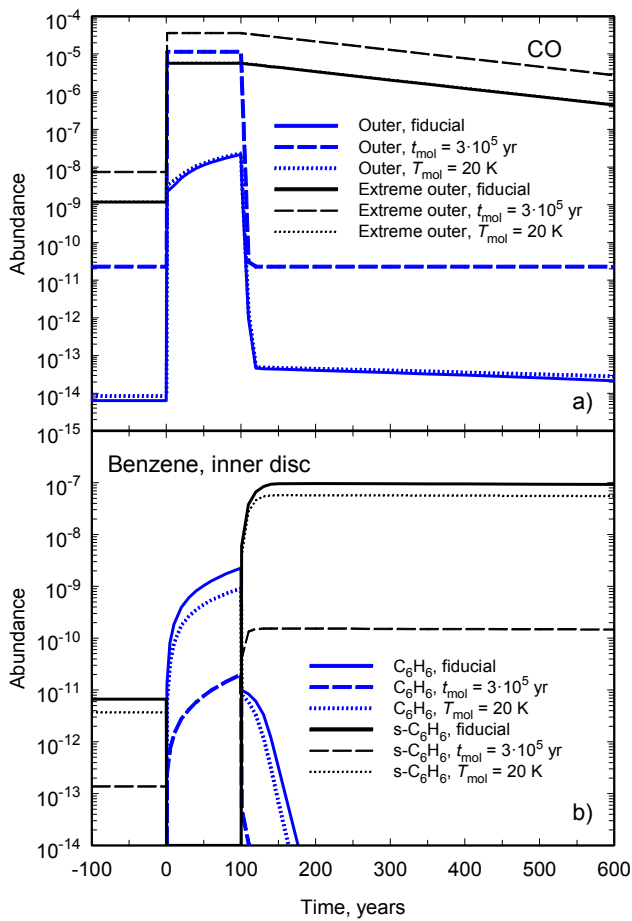}
    \caption{Evolution of CO and benzene abundances in models with different parameters of the molecular cloud stage.}
    \label{fig13}
\end{figure}

In the model with the shorter molecular cloud stage, CO ice is initially more abundant, and this leads to a greater gas-phase CO abundance during the outburst. However, this change is only marginal in the inner and intermediate disc regions as by the end of the pre-outburst stage nearly all carbon atoms are anyway locked in gas-phase CO molecules. The most drastic change is observed in the outer disc region, where the $x_{\rm o}$ value for CO gas raises from $2\times10^{-8}$ in the fiducial model to $10^{-5}$ in the model with the shorter molecular cloud stage (blue dashed line in Fig.~\ref{fig13}a). A similar but a more modest increase is observed in the extreme outer disc region (black dashed line in Fig.~\ref{fig13}a).

However, even minor increase of the CO abundance in the inner disc steals C atoms from other species, decreasing their outburst and post-outburst abundances. Most important for our conclusions is the outburst gas-phase benzene abundance, which is decreased by two orders of magnitude due to shorter molecular cloud stage (blue dashed line in Fig.~\ref{fig13}b). The post-outburst abundance of benzene ice drops even stronger, by three orders of magnitude, down to $10^{-10}$ (black dashed line in Fig.~\ref{fig13}b). Thus, the shorter duration of the molecular cloud stage does not change our overall conclusion on the benzene evolution in the inner disc during and after the outburst, but abundances of this molecule at various stages become much smaller. In the intermediate disc region, the post-outburst amount of benzene ice stay high irrespective of the molecular stage duration. Outburst abundances of some other hydrocarbons in the inner disc, like C$_2$H$_2$ and H$_2$C$_3$O are also decreased in this model. It is noteworthy that methanol itself is not much influenced significantly by this change in the model setup.

A rise of the temperature during the molecular stage plays a much less important role, mostly affecting CO$_2$ ice itself and only in the extreme outer disc region. Some other molecules, also show strong dependence on $T_{\rm mol}$, but at a very low abundance level.

A general conclusion is that while abundances of some molecules change noticeably due to variations in parameters of the molecular cloud stage, almost none of them change the behaviour type. The occurring changes are mostly formal, with molecules shuffling between types 1 and 2. Very few molecules move from the first type to the third type, like acetylene, for example, which has a factor of several lower pre-outburst abundance in the model with the shorter molecular cloud stage.

As was mentioned earlier, we have also tried to match the model and observed initial ice abundances by altering the chemical network. Specifically, we have added the hydrogen abstraction reaction s-CH$_3$OH~+~s-H $\rightarrow$~s-CH$_2$OH~+~s-H$_2$. Its high activation energy prevents it from producing any significant changes both in the molecular cloud stage and in the disc stage. However, we can make it more important by artificially reducing its activation energy down to, say, 1500\,K. While this approach is definitely non-physical, it allows revealing possible long-lasting effects of the model alteration aimed to achieving different initial abundances.

CO behaviour in this model is nearly identical to its behaviour in the model with the shorter molecular cloud stage. Quite expectedly, less effective methanol synthesis increases abundances of other hydrocarbons, which compete with methanol for carbon atoms. These changes are mostly bound to the intermediate disc region, where outburst abundances of CH$_3$OCH$_3$, HCOOCH$_3$, and C$_2$H$_5$OH rise above $10^{-9}$, being less than $10^{-11}$ in the fiducial model. One non-trivial outcome of the network change is that methanol gas-phase abundance actually {\em increases} during the pre-outburst stage in the outer and extreme outer disc regions. This is the methanol that is released to the gas phase (where it is not affected by the backward reaction) from dust surfaces due to reactive desorption. In general, in this case we also conclude that while the introduced change in the network affects some abundances, it does not alter the species behaviour. All the variations in the species distribution over various types are formal and mostly affect compounds with very small abundances.

In the end of this subsection, we mention two other ways to control the methanol formation efficiency. One of them is to vary activation barriers for hydrogen addition reactions s-H~+~s-CO and s-H~+~s-H$_2$CO, which are links in the reaction chain leading to methanol formation. In the current study, we adopt 2500\,K activation energies for both reactions \citep{Sosa,woon}, but these energies can be different depending on the mantle composition, structure and/or temperature \citep{watanabe2003,korchagina2017}. Various experimental and theoretical studies tend to produce controversial results on activation barriers for these reactions, and also the role of tunnelling at different dust temperatures is not clear \citep{watanabe2003,fuchs,ANDERSSON201131}. However, most of these results seem to indicate higher hydrogen addition rates than those adopted by us, for example, due to effects of nearby mantle molecules \citep{woon,2014A&A...572A..70R}. Thus, taking them into account would increase overall methanol ice production.

On the other hand, in our model we use a two-phase (gas+dust) formalism, which tend to overestimate the efficiency of the surface CH$_3$OH synthesis \citep[e.g.,][]{2012A&A...538A..42T}. An alternative would be to use a multi-layer approach that accounts for CO trapping within the mantle making it unavailable for further processing. Both options deserve separate studies.

\section{Discussion}

% HCN CH3CN CH2CN H2CO CH3OH

In this study we present a detailed analysis of chemical processes, which may occur in a young protoplanetary disc, exposed to episodic luminosity outbursts. The outburst outcomes can be generally classified into several types. The zero type means that the abundance of a certain species changes due to the outburst by less than an order of magnitude. The only species that demonstrates the behaviour of the zero type everywhere in the disc is helium. The first type implies rapid growth or fall-off of the abundance at the start of the outburst, no evolution during the outburst, and rapid restoration of the pre-outburst abundance after the end of the outburst. This type is usually related to evaporation and subsequent re-freeze out, typical for species with large desorption energies and/or species, which form on dust surfaces. Species that belong to the first type in the fiducial model may demonstrate the different response type in models with increased $a_{\max}$. Due to the adopted treatment of transition from molecular cloud stage to the quiescent disc stage we force these species to stay on dust surfaces, even if the total surface area shrinks significantly. The ice content is reset to the new equilibrium state after the temperature returns to its pre-outburst value. In this state, species that had previously been locked in icy mantles may stay in the gas-phase for a longer time.

The second type of behaviour is shown by the species that are synthesized or destroyed during the outburst thanks to elevated temperature. The significance of this type is that it may lay the ground for some post-outburst abundance changes, for example, for surface counterparts of the gas-phase species belonging to this type. Specifically, the gas-phase products of the outburst-triggered chemistry may freeze-out after the end of the outburst, with their post-outburst surface abundances exceeding significantly the pre-outburst values.

The third type of behaviour is demonstrated by species that are preserved or even synthesized at the post-outburst stage from compounds, having been released due to the outburst-triggered chemistry. Some complex organic molecules, related to this type, are promising outburst tracers as their pre-outburst abundances are negligibly small. This means that a mere presence of these molecules in the disc is a signature of recent accretion activity. Inspection of Tables~\ref{ddgr3} and~\ref{agr3} shows that most promising outburst tracers in the star vicinity are organic ices, like solid-phase benzene, propionitrile, methylamine, which are produced due to a post-outburst evolution of the type~2 species (see above).  While these species are not directly observable (in the absence of a suitable background source, illuminating the disc from behind), a subsequent outburst may desorb them, thus revealing their presence. Direct observations of these species may be possible with the JWST in edge-on discs. Their production in the outburst(s) may have also played a role in the early evolution of the Solar System.

Organics are also sensitive long-lasting tracers of outburst chemistry farther out from the star. Two remarks should be made. First, sensitive species mostly reside  at the extreme outer disc (where physical conditions are actually similar to those in inner envelopes of Class~I systems). So, their ability to trace past outbursts may prove useless in compact discs with sizes of the order of 100~au or less. Second, at the extreme outer disc sensitivity of some of the molecules listed in Table~\ref{ddgr3} weakens as dust grains grow. At the same time, their `sensitivity zone' shifts toward the star (see Table~\ref{nowyouseemenowyoudont}). On the one hand, this implies that these molecules become outburst tracers even in compact discs as dust evolves in them. On the other hand, in more extended discs with evolved dust abundances of these species enhanced by the outburst may be lost in comparison to the chemically inert extreme outer region. Thus, we should be mostly interested in species that remain sensitive to the outburst irrespective to the maximum grain size. Some of them are listed in Table~\ref{finalresult} that includes species having $x_{\rm p}>10^{-12}$ at any $a_{\max}$.

\begin{table}
\caption{Species sensitive to the past luminosity outburst irrespective to the maximum grain size. {Values of $x_{\rm}$ and $R_{\rm pq}$ are given for $a_{\max}$ indicated in subheadings.}}
\label{finalresult}
\begin{tabular}{llllll}
\hline
 & & & & & Loca- \\
Species & \multicolumn{2}{c}{$x_{\rm p}$} & \multicolumn{2}{c}{$R_{\rm pq}$} & tion \\
\cline{2-3}\cline{4-5}
        & $10^{-4}$ cm & 10 cm & $10^{-4}$ cm & 10 cm &  \\
\hline
C             & 1.3(--12) & 2.8(--12) & 1.6(02) & 1.1(02) & I  \\
C$_4$S        & 5.7(--12) & 1.1(--11) & 6.6(02) & 1.0(03) & I  \\
s-C$_4$S      & 5.5(--10) & 3.5(--12) & 6.6(02) & 1.0(03) & I  \\
C$_6$H$_6$    & 9.3(--08) & 2.0(--08) & 1.4(04) & 1.7(01) & I  \\
s-CH$_3$CN    & 3.6(--11) & 4.2(--12) & 2.2(02) & 8.9(01) & I  \\
s-CH$_3$OH    & 2.1(--12) & 1.5(--11) & 1.9(03) & 1.6(02) & I  \\
CS            & 1.3(--11) & 1.3(--10) & 4.2(01) & 9.1(01) & I  \\
H$_2$CS       & 2.4(--10) & 1.3(--09) & 2.7(01) & 4.9(01) & I  \\
s-H$_3$C$_5$N & 1.2(--09) & 1.3(--09) & 2.5(02) & 8.3(01) & I  \\
s-H$_5$C$_3$N & 9.0(--09) & 1.6(--09) & 8.6(02) & 1.5(01) & I  \\
HCN           & 2.4(--11) & 5.6(--11) & 6.2(01) & 2.5(01) & I  \\
HCP           & 4.7(--12) & 2.4(--12) & 1.3(02) & 2.7(01) & I  \\
s-HCSi        & 3.0(--11) & 1.3(--11) & 3.8(02) & 5.8(01) & I  \\
P             & 6.9(--11) & 7.0(--09) & 4.3(02) & 1.7(02) & I  \\
SiC$_2$H      & 1.6(--10) & 5.3(--07) & 2.2(01) & 2.9(06) & I  \\
\hline
s-C$_6$H$_6$  & 4.0(--07) & 6.0(--11) & 8.3(01) & 4.9(02) & IM \\
\hline
s-HCOOCH$_3$  & 6.8(--10) & 1.7(--11) & 7.4(03) & 4.9(15) & O  \\
s-NH$_2$OH    & 7.6(--10) & 1.0(--11) & 2.3(15) & 1.2(02) & O  \\
\hline
C$_2$H$_3$    & 8.2(--11) & 3.6(--10) & 2.0(01) & 1.1(03) & EO \\
C$_3$H$_3$    & 2.1(--11) & 3.6(--12) & 1.3(01) & 3.9(01) & EO \\
C$_3$H$_5^+$  & 1.2(--11) & 1.1(--11) & 3.3(01) & 3.8(01) & EO \\
CH$_3$CHO     & 1.3(--12) & 5.5(--12) & 2.5(04) & 4.6(01) & EO \\
CH$_3$CN      & 1.4(--11) & 2.8(--10) & 3.5(03) & 1.2(01) & EO \\
CH$_4$        & 9.5(--07) & 3.0(--05) & 1.1(03) & 2.9(02) & EO \\
Cl            & 9.8(--11) & 9.6(--10) & 1.2(03) & 3.9(01) & EO \\
HCl           & 2.0(--12) & 3.5(--11) & 1.5(04) & 7.7(01) & EO \\
NS            & 4.8(--11) & 2.8(--10) & 1.3(03) & 3.4(01) & EO \\
P             & 1.6(--11) & 1.8(--10) & 8.2(04) & 5.1(03) & EO \\
\hline
\end{tabular}
\end{table}

There are a few molecules that show a long-lasting memory of the outburst {\em both\/} in the dark disc and the disc atmosphere. The most noticeable of them is hydroxylamine (NH$_2$OH) that is extremely sensitive to the outburst either in the gas phase or in the solid phase. Its recent non-detection in the low-mass protostellar binary IRAS 16293--2422 \citep{2018arXiv180800251L} is circumstantially consistent with our result and could make hydroxylamine especially valuable target for future observations. Also we note that the conclusion of \citet{2018arXiv180800251L} on the limited importance of hydroxylamine for the formation of amino acids can be relaxed in the discs experiencing outbursts. On the other hand, the chemistry of this molecule is very uncertain, and the corresponding conclusions should be treated with extreme caution.

% Ratios s-CO/s-CO$_2$, N$_2$H$^+$/CO, HCO$^+$/CO, H$_2$CO/CO, HCO$^+$/H$_2$O

Apart from affecting specific molecular abundances, the outburst effect can be seen in varying C/O ratios in gas and solid phases (Table~\ref{coratio}). The total (gas+mantle) ratio of C and O abundances in the adopted element set is 0.41; after the end of the molecular cloud stage the C/O ratio is 0.62 for the gas phase and 0.39 for the solid phase. As in the pre-outburst stage abundant species are frozen-out, the solid-phase C/O value coincides with the global value almost everywhere in the dark disc except for its inner region. The gas phase is relatively richer in carbon than the solid phase in the cold dark disc regions, however, this is because all the major carbon and oxygen compounds are frozen out and the C/O ratio is determined by less abundant molecules, like CH$_4$ and HNO. In the disc atmosphere, the pre-outburst gas-phase C/O ratio is close to one everywhere except for the inner region due to dominance of CO. Farther out in the disc CO is mixed with gas-phase methane, which make the C/O ratio greater than one.

In the inner and intermediate dark disc regions and everywhere in the disc atmosphere, the outburst causes complete mantle evaporation, so that the gas-phase C/O ratio at this stage is equal to the global ratio of 0.41. The solid-phase C/O varies significantly over the model disc, however these variations are caused by less abundant species. The dramatic solid-phase C/O increase in the intermediate dark disc during the outburst is caused by complex hydrocarbons, predominantly benzene and s-C$_9$H$_2$, which stay on dust surfaces even during the outburst, when most O-bearing species are evaporated from icy mantles. After the outburst, the C/O ratios in different disc regions rapidly (within a few centuries) return to their quiescent values, except for the outer disc atmosphere. There the gas phase is richer in carbon than in oxygen prior to the outburst due to high methane abundance. Methane is destroyed during the outburst and does not reform after its end. Surface hydrocarbons (first of all, methanol ice and ethane ice) that are quite abundant at the quiescent stage are completely evaporated and destroyed by the UV radiation during the outburst and do not reform at the post-outburst stage, causing significant decrease of the solid-state C/O ratio.

\begin{table*}
\caption{C/O ratios in various disc regions and at various evolutionary stages. A slash separates ratios for gas and solid phases. The total C/O ratio is 0.41.}
\label{coratio}
\begin{tabular}{lllllll}
\hline
Location     &\multicolumn{3}{c}{Dark disc}& \multicolumn{3}{c}{Disc atmosphere} \\
\cline{2-7}
 & Pre-outburst & Outburst & $10^4$ yr after  & Pre-outburst & Outburst & $10^4$ yr after \\
 &  & & the outburst &  &  & the outburst \\
\hline
Inner        & 0.77/0.08 & 0.41/0.16 & 0.85/0.07  & 0.41/0.002 & 0.41/0.41      & 0.41/0.002 \\
Intermediate & 1.06/0.41 & 0.41/272  & 1.06/0.41  & 1.00/0.14      & 0.41/1.60      & 1.04/0.06 \\
Outer        & 2.10/0.41 & 1.55/0.14 & 2.10/0.41  & 1.26/0.31      & 0.41/3.52      & 0.56/0.003 \\
Extreme outer& 1.46/0.41 & 5.37/0.22 & 1.46/0.41  & 1.49/0.37      & 0.43/0.01  & 2.03/0.03 \\
\hline
\end{tabular}
\end{table*}

Both during and after the outburst abundances of complex and even some simple species are controlled by intricate combination of processes. This raises an obvious question: to what degree are our results model-dependent? For example, some key processes in our study proceed on dust surfaces and thus sensitively depend on the adopted desorption energies, which determine mantle evaporation during the outburst and surface species mobility. The choice of reaction themselves defines operating pathways and, in particular, causes accumulation of species, which represent dead ends in the adopted reaction network.

As the chemical evolution during and after outburst is tightly bound to mantle evaporation and re-adsorption, it sensitively depends on the adopted set of desorption temperatures. Some additional outburst tracers may appear if we take into account the dependence of desorption energies on the mantle composition. For example, with the low CH$_3$OH desorption energy (2060\,K) from \cite{1993MNRAS.261...83H}, methanol would be a much prominent outburst tracer than it is currently, with the desorption energy of 5530\,K. However, the importance of methanol as an outburst tracer may rise if we take into account its co-desorption with other molecules \citep[e.g.][]{2014A&A...564A...8M,2018arXiv180104846L}.

A very important element of the model is also the presence of a pre-disc chemical evolution \citep[see also][]{2016MNRAS.462..977D,2017arXiv170907863E}. During this stage, an initial molecular inventory is established, which later determines the chemical evolution of the disc material, especially regarding distribution of molecules between solid phase and gas-phase. If chemistry is reset during the early disc build-up, i.e. molecular cloud chemical inventory is wiped out, our results for the inner disc regions would be different. Farther out from the star conditions in the disc are more similar to the conditions in the molecular cloud, and the molecular stage becomes less critical.

Current study is focused on the detailed outburst-induced chemistry in some selected disc locations. This allows us to investigate the sensitivity of the results to the different parameters like grain size distribution, outburst duration, and strength. A spatially resolved chemical modelling is the next necessary step to understand chemical pathways of complex molecules in FU Ori-like vigorous environments \citep{2018ApJ...866...46M}.

\section{Conclusions}

In this study, we consider the chemical response of the material of the protoplanetary disc to the luminosity outburst. Three main kinds of behaviour can be identified. In the simplest case the major factor in changing the disc chemical composition is icy mantle evaporation and subsequent re-adsorption. This behaviour is typical for species with large desorption energies and/or species, which form on dust surfaces.

The second kind of evolution is observed for species that are synthesized or destroyed during the outburst thanks to elevated gas and dust temperature. One of the interesting examples of this behaviour in the inner and intermediate disc regions is accumulation of benzene ice that is preserved for many centuries after the outburst. While this excess benzene ice is not directly observable, it may be revealed by subsequent luminosity outbursts. In the inner disc, the amount of the accumulated benzene ice depends on the adopted duration of the molecular cloud stage.

The third type comprises species that are synthesized at the post-outburst stage. Their appearance is related to the outburst chemistry, which enriches the medium with compounds that were absent at the pre-outburst stage. Some complex organic molecules experiencing response of this type, like methyl formate (HCOOCH$_3$), ethenone (CH$_2$CO), and acetonitrile (CH$_3$CN), are promising tracers of past outbursts, especially in the cases, when their pre-outburst abundances are negligibly small. Hydroxylamine is the most prominent species that is influenced by the outburst both in the dark disc and in the atmosphere. It is interesting that neither CO, nor CO$_2$ are good outburst tracers {\it in the disc}. Their abundances change during the outburst, but quickly return to the pre-outburst stage everywhere in the disc, except, perhaps, for its outermost, low-density parts. Nevertheless, the abundance of these species {\it in the envelope} can efficiently be used to trace the past outbursts \citep[e.g.,][]{Vorobyov2013,2015A&A...579A..23J,2017A&A...604A..15R}. The outburst also changes the C/O ratio significantly, but the ratio returns to the pre-outburst value almost everywhere in the disc.

Generally, we conclude that the primary effect of the outburst onto the disc molecular content is the rise in disc temperature and to a much lesser extent the rise in radiation intensity. Increased luminosity heats up the disc material deep in its interior and triggers some chemical processes that do not occur at all or proceed very slowly at the pre-outburst quiescent stage. An important feature of these processes is that they produce enhanced surface or gas-phase abundances of some species during the outburst that are preserved well after the outburst. This is especially true for organic species and for species with less abundant elements, like Si or P. While our conclusions are model-dependent to a certain degree, we show that they can actually be explained in basic terms and are thus quite reliable.

\section*{Acknowledgements}

We are grateful to an anonymous referee for valuable suggestions that have helped a lot to clarify our presentation. The study was supported by the RFBR grant 17-02-00644.

\bibliographystyle{mnras}
\bibliography{wiebe}

\appendix

\section{Desorption energies and sublimation temperatures}

\begin{figure*}
\includegraphics[width=0.3\textwidth]{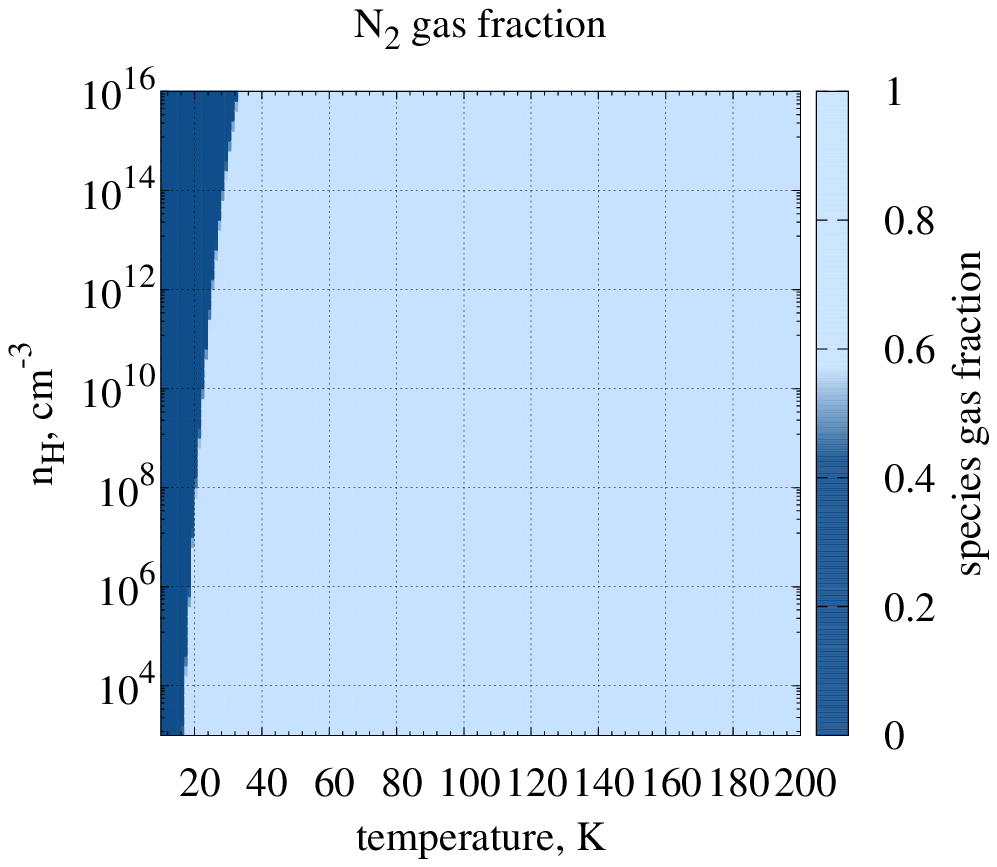}\hspace{5mm}
\includegraphics[width=0.3\textwidth]{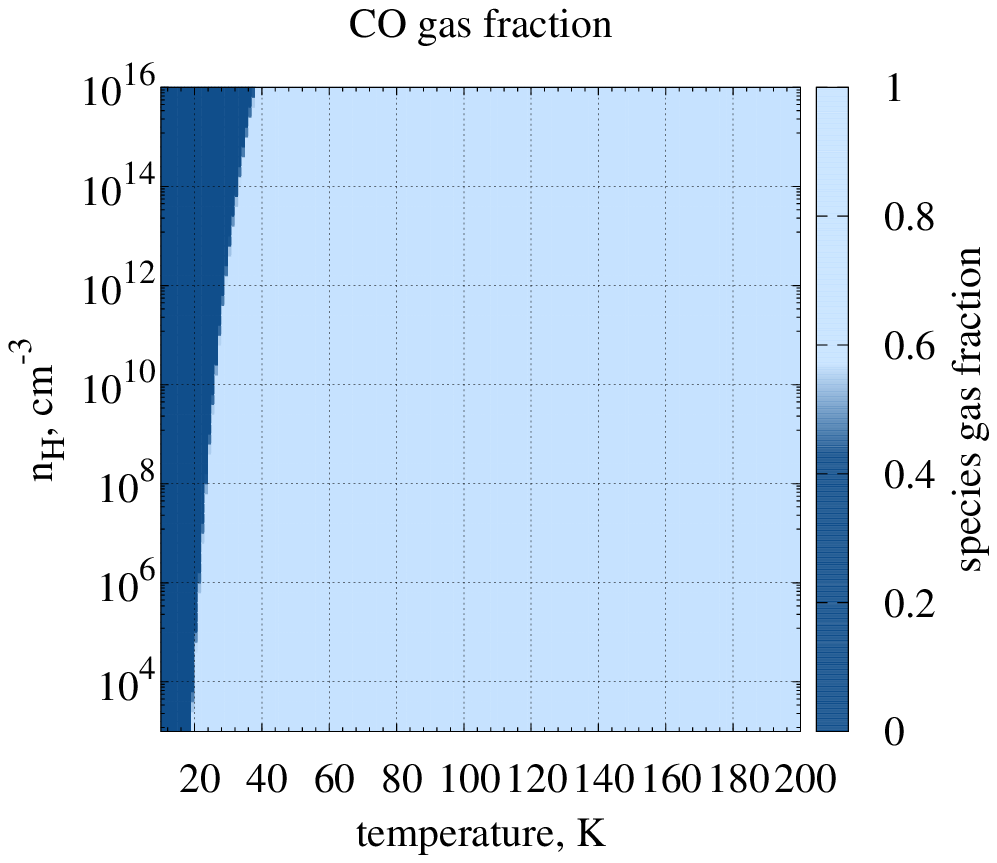}\\
\includegraphics[width=0.3\textwidth]{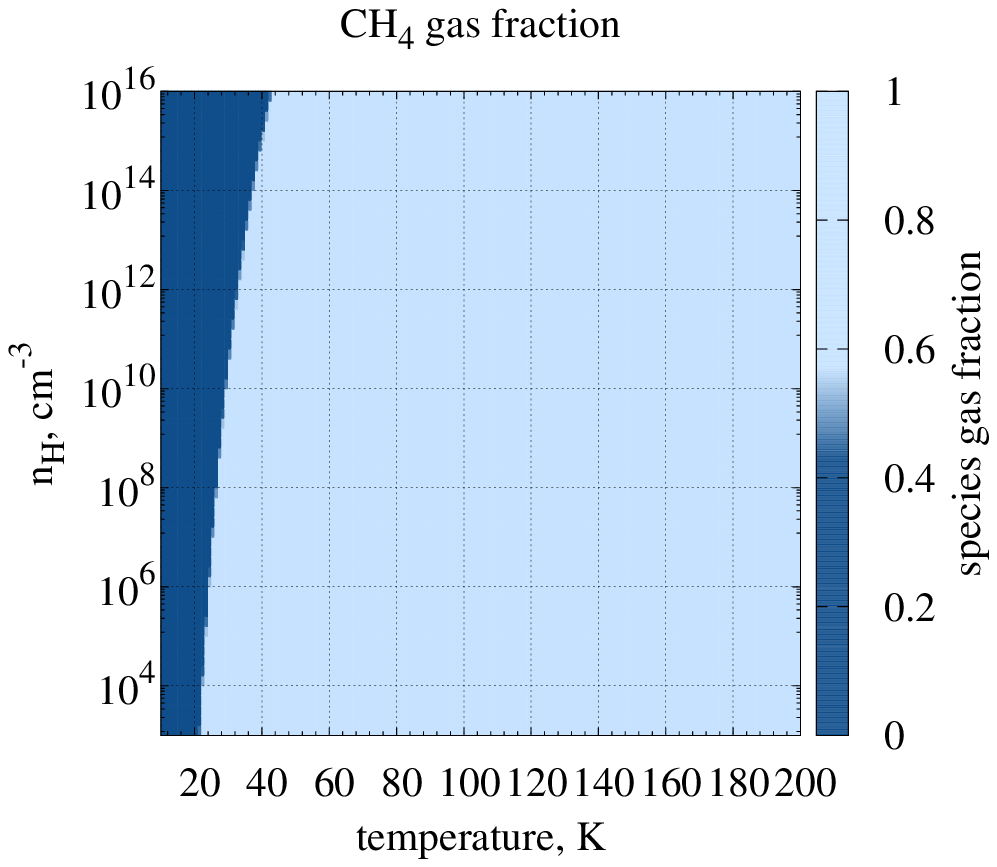}\hspace{5mm}
\includegraphics[width=0.3\textwidth]{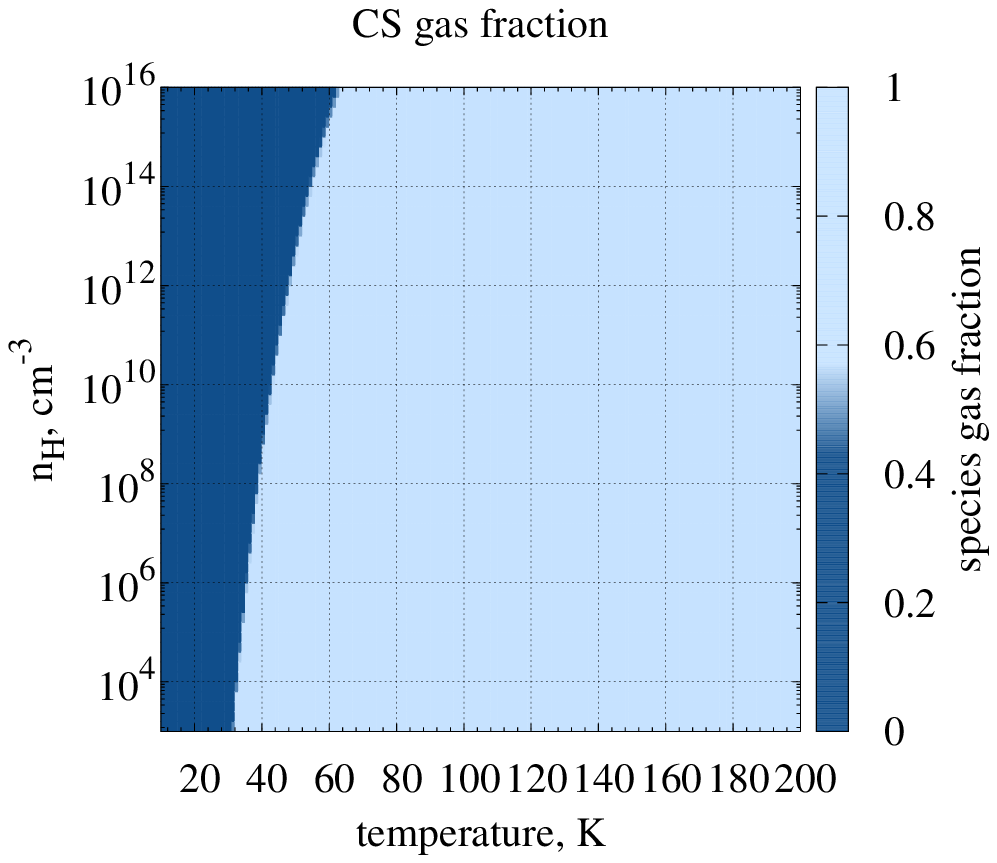}\\
\includegraphics[width=0.3\textwidth]{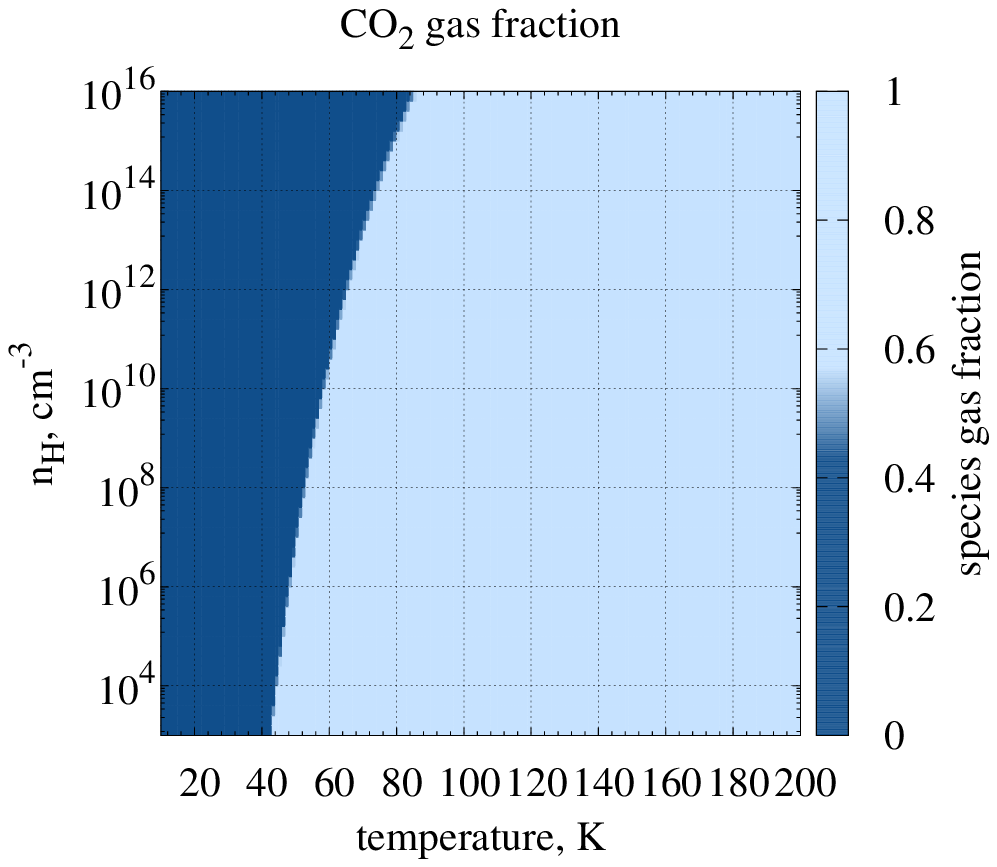}\hspace{5mm}
\includegraphics[width=0.3\textwidth]{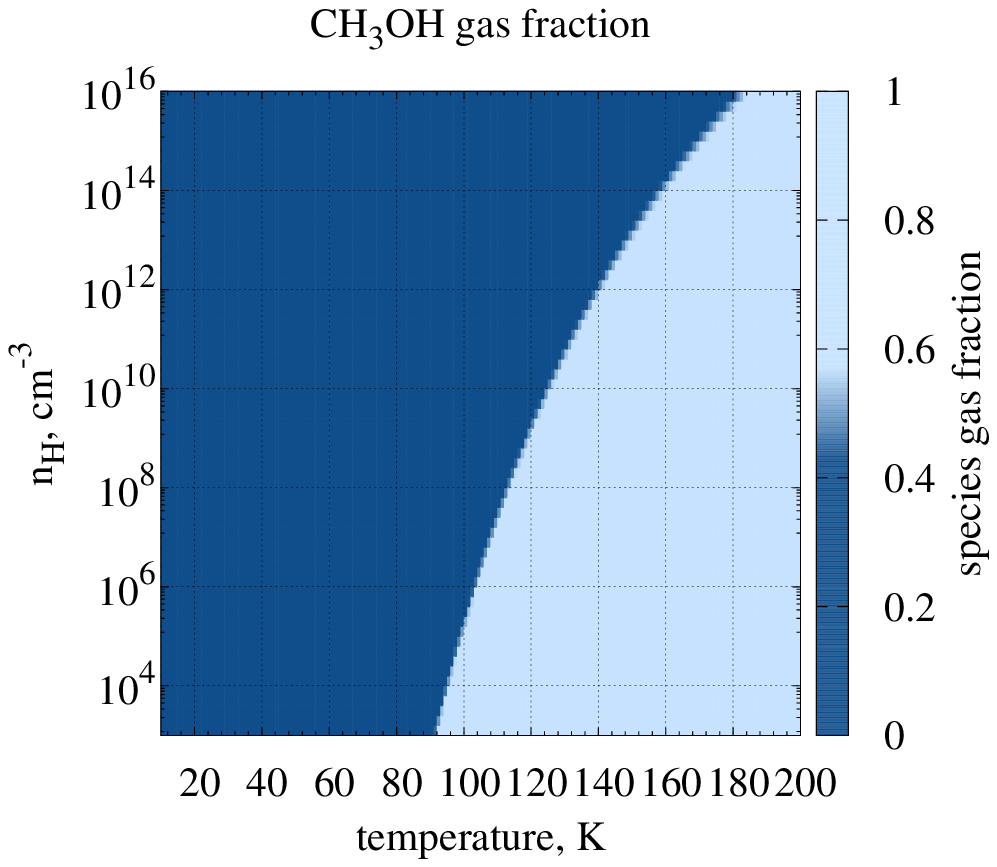}\\
\includegraphics[width=0.3\textwidth]{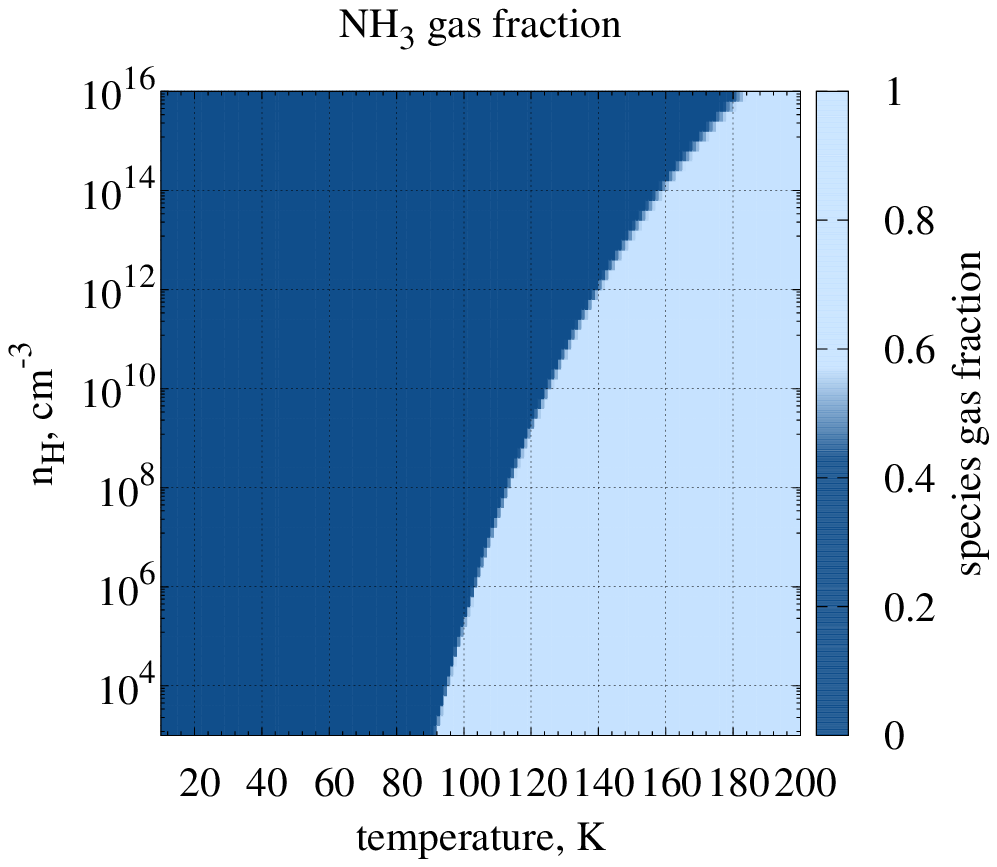}\hspace{5mm}
\includegraphics[width=0.3\textwidth]{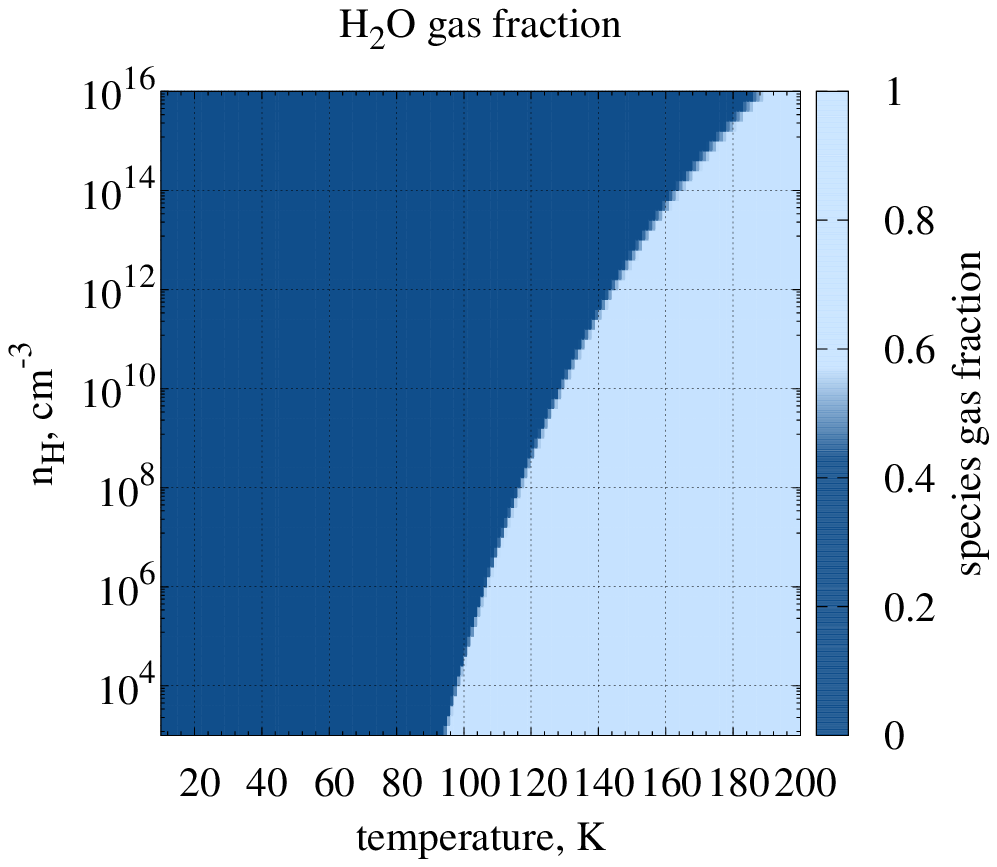}
\caption{Steady-state gas-phase fractions of some molecules as functions of density and temperature.}
\label{apfig}
\end{figure*}

In Fig.~\ref{apfig} we present how steady-state gas-phase fractions of some molecules depend on density and temperature for the adopted disc model. As the transition between a region where a certain molecule is totally frozen out and a region where it is mostly contained in the gas-phase is in all cases quite sharp, these diagrams actually show the dependence of a sublimation temperature on the physical conditions. To determine sublimation temperatures of volatile species, we follow \citet{2015A&A...582A..41H}. The ratio between the species number density in ice and in gas was calculated using their Eqs.~7--8 with similar assumptions on the grain size (0.1\,$\mu$m), dust number density $n_{\rm d}=10^{-12}n_{\rm H}$, but for our assumed site density $N_{\rm ss}=4\cdot10^{14}$\,cm$^{-2}$. Adopted desorption energies are listed in Table~\ref{desen}.

\begin{table*}
\caption{Desorption energies used in the paper.}
\label{desen}
\begin{tabular}{llllllll}
\hline
Species & $T_{\rm des}$, K & Species & $T_{\rm des}$, K & Species & $T_{\rm des}$, K & Species & $T_{\rm des}$, K \\
\hline
C             &   800   &  C$_8$H$_2$     &       7390 &  H$_2$S$_2$   &   3100 &  OH            &        2850 \\
C$_{10}$        &  8000   &  C$_8$H$_3$     &       7840 &  H$_3$C$_5$N  &   7080 &  S             &        1100 \\
C$_2$         &  1600   &  C$_8$H$_4$     &       8290 &  H$_3$C$_7$N  &   8680 &  S$_2$            &        2200 \\
C$_2$H        &  2140   &  C$_9$          &       7200 &  H$_3$C$_9$N  &  10300 &  Si            &        2700 \\
C$_2$H$_2$    &  2590   &  C$_9$H         &       7740 &  H$_4$C$_3$N  &   5930 &  SiC           &        3500 \\
C$_2$H$_3$    &  3040   &  C$_9$H$_2$     &       8190 &  H$_5$C$_3$N  &   6380 &  SiH           &        3150 \\
C$_2$H$_4$    &  3490   &  C$_9$H$_3$     &       8640 &  HC$_2$NC     &   4580 &  SiH$_2$          &        3600 \\
C$_2$H$_5$    &  3940   &  C$_9$H$_4$     &       9090 &  HC$_2$O      &   2400 &  SiH$_3$          &        4050 \\
C$_2$H$_5$OH  &  6580   &  C$_9$N         &       8000 &  HC$_3$N      &   4580 &  SiH$_4$          &        4500 \\
C$_2$H$_6$    &  4390   &  CH             &        925 &  HC$_3$O      &   3200 &  SiO           &        3500 \\
C$_2$N        &  2400   &  CH$_2$         &       1050 &  HC$_5$N      &   6180 &  SiS           &        3800 \\
C$_2$O        &  1950   &  CH$_2$CN       &       4230 &  HC$_7$N      &   7780 &  SO            &        2600 \\
C$_2$S        &  2700   &  CH$_2$CO       &       2200 &  HC$_9$N      &   9380 &  SO$_2$           &        3400 \\
C$_3$         &  2400   &  CH$_2$NH       &       3430 &  HCCN         &   3780 &  C$_2$H$_6$CO        &        2820 \\
C$_3$H        &  2940   &  CH$_2$NH$_2$   &       5010 &  HCN          &   2050 &  C$_3$P           &        3250 \\
C$_3$H$_2$    &  3390   &  CH$_2$OH       &       5080 &  HCNC$_2$     &   4580 &  C$_4$P           &        3450 \\
C$_3$H$_3$    &  3840   &  CH$_3$         &       1180 &  HCO          &   1600 &  CCl           &        2500 \\
C$_3$H$_3$N   &  5480   &  CH$_3$C$_3$N   &       6480 &  HCOOCH$_3$   &   6300 &  CCP           &        2760 \\
C$_3$H$_4$    &  4290   &  CH$_3$C$_4$H   &       5890 &  HCOOH        &   5570 &  CH$_2$PH         &        2470 \\
C$_3$N        &  3200   &  CH$_3$C$_5$N   &       7880 &  HCS          &   2350 &  Cl            &        1220 \\
C$_3$O        &  2750   &  CH$_3$C$_6$H   &       7490 &  He           &    100 &  ClO           &        2140 \\
C$_3$S        &  3500   &  CH$_3$C$_7$N   &       9480 &  HNC          &   2050 &  CP            &        2000 \\
C$_4$         &  3200   &  CH$_3$CHO      &       2870 &  HNC$_3$      &   4580 &  H$_2$SiO         &        7000 \\
C$_4$H        &  3740   &  CH$_3$CN       &       4680 &  HNCO         &   2850 &  HCCP          &        1820 \\
C$_4$H$_2$    &  4190   &  CH$_3$NH       &       3550 &  HNO          &   2050 &  HCP           &        2000 \\
C$_4$H$_3$    &  4640   &  CH$_3$OCH$_3$  &       3150 &  HS           &   1450 &  HCSi          &        8600 \\
C$_4$H$_4$    &  5090   &  CH$_3$OH       &       5530 &  HS$_2$       &   2650 &  HNSi          &        8700 \\
C$_4$N        &  4000   &  CH$_4$         &       1300 &  Mg           &   5300 &  HPO           &        2400 \\
C$_4$S        &  4300   &  CH$_5$N        &       6580 &  MgH          &   5750 &  N$_2$O           &        2300 \\
C$_5$         &  4000   &  CHNH           &       3300 &  MgH$_2$      &   6200 &  NH$_2$CN         &        2000 \\
C$_5$H        &  4540   &  CN             &       1600 &  N            &    800 &  NO$_2$           &        1860 \\
C$_5$H$_2$    &  4990   &  CO             &       1150 &  N$_2$        &   1000 &  P             &        1560 \\
C$_5$H$_3$    &  5440   &  CO$_2$         &       2580 &  N$_2$H$_2$   &   4760 &  PH            &        1600 \\
C$_5$H$_4$    &  5890   &  CS             &       1900 &  Na           &  11800 &  PH$_2$           &        1650 \\
C$_5$N        &  4800   &  Fe             &       4200 &  NaH          &  12200 &  PN            &        2250 \\
C$_6$         &  4800   &  FeH            &       4650 &  NaOH         &  14600 &  PO            &        2350 \\
C$_6$H        &  5340   &  H              &        624 &  NH           &   2380 &  SiC$_2$          &        2660 \\
C$_6$H$_2$    &  5790   &  H$_2$          &        552 &  NH$_2$       &   3960 &  SiC$_2$H         &        2750 \\
C$_6$H$_3$    &  6240   &  H$_2$C$_3$N    &       5030 &  NH$_2$CHO    &   5560 &  SiC$_2$H$_2$        &        3000 \\
C$_6$H$_4$    &  6690   &  H$_2$C$_3$O    &       3650 &  NH$_2$OH     &   6810 &  SiC$_3$          &        3200 \\
C$_6$H$_6$    &  7590   &  H$_2$C$_5$N    &       6630 &  NH$_3$       &   5530 &  SiC$_3$H         &        3300 \\
C$_7$         &  5600   &  H$_2$C$_7$N    &       8230 &  NO           &   1600 &  SiC$_4$          &        3800 \\
C$_7$H        &  6140   &  H$_2$C$_9$N    &       9830 &  NS           &   1900 &  SiCH$_2$         &        2100 \\
C$_7$H$_2$    &  6590   &  H$_2$CN        &       2400 &  O            &    800 &  SiCH$_3$         &        2200 \\
C$_7$H$_3$    &  7040   &  H$_2$CO        &       2050 &  O$_2$        &   1000 &  SiN           &        2100 \\
C$_7$H$_4$    &  7490   &  H$_2$CS        &       2700 &  O$_2$H       &   3650 &  SiNC          &        2700 \\
C$_7$N        &  6400   &  H$_2$O         &       5700 &  O$_3$        &   1800 &  SiO$_2$          &        2730 \\
C$_8$         &  6400   &  H$_2$O$_2$     &       5700 &  OCN          &   2400 &                &             \\
C$_8$H        &  6940   &  H$_2$S         &       2740 &  OCS          &   2890 &                &             \\
\hline
\end{tabular}
\end{table*}

% Don't change these lines
\bsp	% typesetting comment
\label{lastpage}
\end{document}